\documentclass[a4paper]{lipics-v2019}
\nolinenumbers 
\hideLIPIcs  

\usepackage{amsfonts}
\usepackage{amsmath}
\usepackage{amssymb}
\usepackage{amsthm}
\usepackage{stmaryrd}

\usepackage{proof} 
\usepackage{qtree} 
\usepackage[all]{xy} 

\usepackage{tikz}
\usetikzlibrary{arrows.meta} 

\usepackage{hyperref} 

\usepackage{lstcoq}
\usepackage{lstocaml}

\newcommand{\IF}{\mbox{ if }}

\newcommand{\OR}{\mbox{ or }}

\newcommand{\B}{{\lambda^{\scriptscriptstyle \rm BDdL}}}

\newcommand{\match}[6]{\mathbf{smatch}\ #1\ \mathbf{return}\ #2\ \mathbf{with}\,[ #3 \Rightarrow #4 \mid #5 \Rightarrow #6 ]}
\newcommand{\llet}[4]{\mathbf{let}\ #1 \of #2 := #3 \ \mathbf{in}\ #4}
\newcommand{\llett}[3]{\mathbf{let}\ #1 := #2 \ \mathbf{in}\ #3}
\newcommand{\coe}{\mathbf{coe}}

\newcommand{\PTS}{\mathbf{LF}} 

\newcommand{\metastrip}[1]{\widehat{#1}}
\newcommand{\uprefine}{\overset{\Uparrow}{\rightsquigarrow}}
\newcommand{\downrefine}{\overset{\Downarrow}{\rightsquigarrow}}
\newcommand{\forcetype}{\overset{\mathcal F}{\rightsquigarrow}}
\newcommand{\upessence}{\overset{\mathcal{E}^\Uparrow}{\rightsquigarrow}}
\newcommand{\downessence}{\overset{\mathcal{E}^\Downarrow}{\rightsquigarrow}}

\newcommand{\unif}{\overset{?}{=}}
\newcommand{\unifresult}{\overset{\mathcal U}{\rightsquigarrow}}

\newcommand{\at}{\,}
\newcommand{\of}{{:}}

\newcommand{\notion}{\mapsto}

\newcommand{\subst}[3]{#1 [ #3 / #2 ]}

\newcommand{\vecsubst}[3]{#1 \overrightarrow{[ #3 / #2 ]}}

\newcommand{\spair}[2]{\langle #1 , #2 \rangle}
\newcommand{\ssum}[2]{[  #1 , #2 ]}

\newcommand{\prl}{{\sf pr}_{\sf 1}\,}
\newcommand{\prr}{{\sf pr}_{\sf 2}\,}
\newcommand{\pri}{{\sf pr}_i\,}

\newcommand{\inl}[1]{{\sf in}^{#1}_{\sf 1}\,}
\newcommand{\inr}[1]{{\sf in}^{#1}_{\sf 2}\,}

\newcommand{\innl}{{\sf in}_{\sf 1}\,}
\newcommand{\innr}{{\sf in}_{\sf 2}\,}
\newcommand{\inni}{{\sf in}_i\,}

\newcommand{\essence}[1]{\mathopen \wr\,#1\,\mathclose \wr}

\newcommand{\Impl}{\supset}

\newcommand{\notvdash}{\not\,\vdash}

\newcommand{\sigmadash}{\vdash_\Sigma}

\newcommand{\DLF}{{\mathrm{LF}_\Delta}}

\newcommand{\Type}{{\mathsf {Type}}}
\newcommand{\Kind}{{\mathsf {Kind}}}

\newcommand {\FV}{{\mathsf {Fv}}}

\newcommand {\loc}{\mbox{@}}

\newcommand{\eqdef}{\stackrel{\mathit{def}}{=}}
\renewcommand{\leq}{\leqslant}

\newcommand{\s}{\sigma}
\renewcommand{\t}{\tau}
\renewcommand{\r}{\rho}
\renewcommand{\omega}{{\tt U}} 

\newcommand{\D}{\Delta}

\newcommand{\adhoc}{\textit{ad hoc}}

\newcommand{\ala}{\textit{\`a la}}

\newcommand{\eg}{\textit{e}.\textit{g}.\ }

\newcommand{\ie}{\textit{i.e.}}

\author{Claude Stolze, \normalsize{IRIF,  University Paris-Diderot, France}}{}{}{}{}

\author{Luigi Liquori, \normalsize{Inria Sophia Antipolis-M\'editerran\'ee, France}}{}{}{}{}
 
\authorrunning{Claude Stolze and Luigi Liquori}
\Copyright{Claude Stolze and Luigi Liquori}

\funding{Work supported by the COST Action CA15123 EUTYPES “The European research network on types for programming and verification''}

\title{A Type Checker for a Logical Framework with Union and Intersection Types}
\titlerunning{A Type Checker for a Logical Framework with Union and Intersection Types}

\begin{document}
\maketitle

\begin{abstract} 
We present the syntax, semantics, and typing rules of {\color{blue} \href{https://github.com/cstolze/Bull}{Bull}}, a prototype theorem prover based on the $\Delta$-Framework, i.e. a fully-typed lambda-calculus decorated with union and intersection types, as described in previous papers by the authors. {\color{blue} \href{https://github.com/cstolze/Bull}{Bull}} also implements a subtyping algorithm for the Type Theory $\Xi$ of Barbanera-Dezani-de'Liguoro. {\color{blue} \href{https://github.com/cstolze/Bull}{Bull}} has a command-line interface where the user can declare axioms, terms, and perform computations and some basic terminal-style features like error pretty-printing, subexpressions highlighting, and file loading. Moreover, it can typecheck a proof or normalize it. These terms can be incomplete, therefore the typechecking algorithm uses unification to try to construct the missing subterms. {\color{blue} \href{https://github.com/cstolze/Bull}{Bull}} uses the syntax of Berardi’s Pure Type Systems to improve the compactness and the modularity of the kernel. Abstract and concrete syntax are mostly aligned and similar to the concrete syntax of Coq. {\color{blue} \href{https://github.com/cstolze/Bull}{Bull}} uses a higher-order unification algorithm for terms, while typechecking and partial type inference are done by a bidirectional refinement algorithm, similar to the one found in Matita and Beluga. The refinement can be split into two parts: the essence refinement and the typing refinement. Binders are implemented using commonly-used de Bruijn indices. We have defined a concrete language syntax that will allow user to write $\Delta$-terms. We have defined the reduction rules and an evaluator. We have implemented from scratch a refiner which does partial typechecking and type reconstruction. We have experimented {\color{blue} \href{https://github.com/cstolze/Bull}{Bull}} with classical examples of the intersection and union literature, such as the ones formalized by Pfenning with his Refinement Types in LF. 
%
%
We hope that this research vein could be useful to experiment, in a proof theoretical setting,  forms of polymorphism alternatives to Girard's parametric one.
\end{abstract}

\vspace{-3mm}
\section{Introduction}
\vspace{-3mm}
This paper provides a unifying framework for two hitherto unreconciled understandings of types: \ie\ types-as-predicates \ala\ Curry and types-as-propositions \ala\ Church. The key to our unification consists in introducing, implementing and experimenting {\em strong proof-functional connectives} \cite{pott80,BDdL,BM94} in a dependent type theory such as the Edinburgh Logical Framework (LF) \cite{LF}.
Both Logical Frameworks and Proof-Functional Logic consider proofs as first-class citizens, albeit differently.

Strong proof-functional connectives take seriously into account the shape of logical proofs, thus allowing for polymorphic features of proofs to be made explicit in formul{\ae}. Hence they provide a finer semantics than classical/intuitionistic connectives, where the meaning of a compound formula depends only on the {\em truth value} or the {\em provability} of its subformul{\ae}. However, existing approaches to strong proof-functional connectives are all quite idiosyncratic in mentioning proofs.
Existing Logical Frameworks, on the other hand, provide a uniform approach to proof terms in object logics, but they do not fully capitalize on subtyping.

This situation calls for a natural combination of the two understandings of types, which should benefit both worlds. On the side of Logical Frameworks, the expressive power of the metalanguage would be enhanced thus allowing for shallower encodings of logics, a more principled use of subtypes \cite{Refine93}, and new possibilities for formal reasoning in existing interactive theorem provers. On the side of type disciplines for programming languages, a principled framework for proofs would be provided, thus supporting a uniform approach to ``proof reuse'' practices based on type theory \cite{ISOSBook,pier91b,caplan,felty94,boite}.

Therefore, in \cite{FSTTCS18} we extended LF with the connectives of {\em strong intersection} (corresponding to  intersection types \cite{BCD,bar2013}) and {\em strong union} (corresponding to union types \cite{macqueen,BDdL}) of Proof-Functional Logic \cite{pott80}. We called this extension the $\Delta$-Framework ($\DLF$), since it builds on the $\Delta$-calculus \cite{festschrift18}. As such, $\DLF$ subsumes many expressive type disciplines in the literature \cite{Refine93,BDdL,BM94,pier91b,caplan}.

It is not immediate to extend the Curry-Howard isomorphism to logics supporting strong proof-functional connectives, since these connectives need to compare the shapes of derivations and do not just take into account the provability of propositions, \ie\ the inhabitation of the corresponding type. In order to capture successfully strong logical connectives such as $\cap$ or $\cup$, we need to be able to express the rules:\\
%
$  \begin{array}{c@{\qquad} c}
    \infer[(\cap I)]{
      A \cap B}{
      {\mathcal D}_1 : A \quad {\mathcal D}_2 : B \quad {\mathcal D}_1 \equiv {\mathcal D}_2}
    &
    \infer[(\cup E)]{
      C}{
      {\mathcal D}_1 : A \supset C \quad {\mathcal D}_2 : B \supset C \quad A \cup B \quad {\mathcal D}_1 \equiv {\mathcal D}_2 }
  \end{array}$\\
%
where $\equiv$ is a suitable equivalence between logical proofs.
Notice that the above rules suggest immediately intriguing applications in polymorphic constructions, \ie\ the \emph{same evidence} can be used as a proof for different statements.

Pottinger \cite{pott80} was the first to study the strong connective $\cap$. He contrasted it to the intuitionistic connective $\wedge$ as follows: \emph{``The intuitive meaning of $\cap$ can be explained by saying that to assert $A \cap B$ is to assert that one has a reason for asserting $A$ which is also a reason for asserting $B$ [while] to assert $A \wedge B$ is to assert that one has a pair of reasons, the first of which is a reason for asserting $A$ and the second of which is a reason for asserting $B$''}.

A logical theorem involving intuitionistic conjunction which does not hold for strong conjunction is $(A\supset A) \wedge (A \supset B \supset A)$, otherwise there should exist a closed $\lambda$-term having simultaneously both one and two abstractions.
L\'opez-Escobar \cite{Lopez-Escobar85} and Mints \cite{Mints89} investigated extensively logics featuring both strong and intuitionistic connectives especially in the context of {\em realizability} interpretations.

Dually, it is in the $\cup$-elimination rule that proof equality needs to be checked. Following Pottinger, we could say that {\em asserting $(A \cup B) \Impl C$ is to assert that one has a reason for $(A \cup B) \Impl C$, which is also a reason to assert $A \Impl C$ and $B \Impl C$.} The two connectives differ since the intuitionistic theorem $((A \supset B) \vee B) \supset A \supset B$ is not derivable for $\cup$, otherwise there would exist a term which behaves both as {\bf I} and as {\bf K}.

Strong connectives arise naturally in investigating the propositions-as-types analogy for intersection and union type assignment systems. From a logical point of view, there are many proposals \cite{Mints89,Refine93,venneri94,roncrove01,miquel01,CLV,BVB,pimronrov12,Ehrhard20,urzy-dude20} to find a suitable logic to fit intersection (and union) : we also refers to \cite{APLAS16,FSTTCS18,stolzephd} for a detailed discussion among logics for intersection and unions.

The $\DLF$ Logical Framework introduced in \cite{FSTTCS18} extends \cite{festschrift18} with union types and
dependent types. The novelty of $\DLF$ in the context of Logical Frameworks, lies in the full-fledged use of strong proof-functional connectives, which to our knowledge has never been explored before. Clearly, all $\Delta$-terms have a computational counterpart.

This paper presents the implementation of {\color{blue} \href{https://github.com/cstolze/Bull}{Bull}} \cite{Bull,stolzephd}, an Interactive Theorem Prover (ITP) based on the $\Delta$-Framework \cite{LFMTP17,FSTTCS18}. The first author wrote this theorem prover from scratch for three years. 
{\color{blue} \href{https://github.com/cstolze/Bull}{Bull}} have a command-line interface program where the user can declare axioms, terms, and perform computations. These terms can be incomplete, therefore the typechecking algorithm uses unification to try to construct the missing subterms.

We have designed and implemented a novel subtyping algorithm \cite{TTCS17} which extends the well-known algorithm for intersection types, designed by Hindley \cite{Hindley82}, with union types. Our subtyping algorithm has been mechanically proved correct in Coq and extracted in Caml, extending the mechanized proof of a subtyping algorithm for intersection types of Bessai \cite{bessai16}.

We have implemented several features. A Read-Eval-Print-Loop allows to define axioms and definitions, and performs some basic terminal-style features like error pretty-printing, subexpressions highlighting, and file loading. Moreover, it can typecheck a proof or normalize it. We use the Berardi's syntax of Pure Type Systems \cite{Berardi90} to improve the compactness and the modularity of the kernel. Abstract and concrete syntax are mostly aligned: the concrete syntax is similar to the concrete syntax of Coq.

We have designed a \textit{higher-order unification algorithm} for terms, while typechecking and partial type inference are done by our \textit{bidirectional refinement algorithm}, similar to the one found in Matita \cite{Refine12}. The refinement can be split into two parts: the essence refinement and the typing refinement.
The bidirectional refinement algorithm aims to have partial type inference, and to give as much information as possible to the unifier. For instance, if we want to find a $?y$ such that $\sigmadash \spair{\lambda x\of\s.x}{\lambda x\of\t.?y} : (\s \rightarrow \s) \cap (\t \rightarrow \t)$, we can infer that $x\of\t\vdash ?y : \t$ and that $\essence{?y} =_\beta x$.

This paper is organized as follows: in Section \ref{sec:lang}, we introduce the language we have implemented. In Section \ref{sec:eval}, we define the reduction rules and explain the evaluation process. In Section \ref{sec:sub}, we present the subtyping algorithm. In Section \ref{sec:unif}, we present the unifier. In Section \ref{sec:refine}, we present the refiner which does partial typechecking and type reconstruction. In Section \ref{sec:repl}, we present the Read-Eval-Print-Loop. In Section \ref{sec:afterbull}, we present possible enhancements of the type theory and of the ITP.
Appendix (at the discretion of the reviewer) contains many interesting encodings that could be typechecked in  {\color{blue} \href{https://github.com/cstolze/Bull}{Bull}} and could help the reviewer to understand the usefulness of adding \adhoc\ polymorphism and proof-functional operators to LF. 

\vspace{-3mm}
\section{Syntax of terms}\label{sec:lang}
\vspace{-3mm}
The syntax for the logical framework we have designed and implemented is as follows:
\begin{displaymath}
\hspace{-1cm}
  \begin{array}{lllllll}
    \Delta,\sigma & ::= & 
                         s,c,v,\_ & \mbox{Sorts, Constants, Variables and Placeholders} \\[1mm]
           & \mid & ?x[\Delta;...;\Delta] & \mbox{Meta-variable} \\[1mm]
           & \mid & \llet{x}{\sigma}{\Delta}{\Delta} & \mbox{Local definition} \\[1mm]
           & \mid & \Pi x\of\sigma.\Delta \ , \ 
            \lambda x\of\sigma.\Delta \ , \  \Delta\at S 
            & \mbox{$\Pi$-abstraction and  $\lambda$-abstraction and application} \\[1mm]
           & \mid & \sigma \cap \sigma \ , \  \sigma \cup \sigma & \mbox{Intersection and Union types} \\[1mm]
           & \mid & \spair{\Delta}{\Delta} \ , \ \prl \Delta \ , \ \prr \Delta & \mbox{Strong pair and Left/Right projections } \\[1mm]
           & \mid & 
           \match{\Delta}{\sigma}{x\of\s}{\Delta}{x\of\s}{\Delta} 
           & \mbox{Strong sum}\\[1mm]
           & \mid & \innl \sigma\at \Delta \ , \ \innr \sigma\at \Delta \ , \  
           \mathbf{coe}\at \sigma\at \Delta & \mbox{Left/Right injections and Coercions} \\[1mm]
           S &::= & () \mid (S{;}\D) & \mbox{Typed Spines}
  \end{array}
\end{displaymath}
By using a Pure Type System approach \cite{Berardi90}, all the terms are read through the same parser. The main differences with the $\Delta$-Framework \cite{FSTTCS18} are the additions of a placeholder and meta-variables, used by the refiner. We also added a $\mathbf{let}$ operator and changed the syntax of the strong sum so it looks more like the concrete syntax used in the implementation. A meta-variable $?x[\Delta_1;...;\Delta_n]$ has the, so called, \emph{suspended substitutions} $\Delta_1;...;\Delta_n$, which will be explained clearly in Subsection \ref{ssec:subst}.
Finally, following the Cervesato-Pfenning jargon \cite{spine}, applications are in \emph{spine form}, \ie\ the arguments of a function are stored together in a list, exposing the head of the term separately.
We also implemented a corresponding syntax for the untyped counterpart of the framework, called \emph{essence} \cite{festschrift18}, where all pure $\lambda$-terms $M$ and spines are defined as follows:
\begin{displaymath}
  \hspace{-1cm}
  \begin{array}{llll}
    M,\varsigma & ::= & s\ , \ c \ , x \ ,  \_  \ , \ ?x[M;...;M]  & \mbox{Sorts, Constants, Variables, Placeholders and Metavariables} \\[1mm]
           & \mid & \llett{x}{M}{M} \ , \ \Pi x\of\varsigma. \varsigma & \mbox{Local definition} \\[1mm]
           & \mid & \Pi x\of\varsigma. \varsigma \ , \  \lambda x.M \ , \ M\at R 
           & \mbox{Dependent product, $\lambda$-abstraction and application} \\[1mm]
           & \mid & \varsigma \cap \varsigma \ , \ \varsigma \cup \varsigma & \mbox{Intersection and Union types } \\[1mm]
            R & ::= & () \mid (R{;}M) & \mbox{Untyped Spines}
  \end{array}
\end{displaymath}
\noindent Note that essences of types (noted $\varsigma$) belongs to the same syntactical set as essences of terms.

\vspace{-3mm}
\subsection{Concrete syntax}
\vspace{-3mm}
The concrete syntax of the terms has been implemented with OCamllex and OCamlyacc. Its simplified syntax is as follows:
{\small
\begin{verbatim}
term ::=
| Type
| let ID [args] [: term] := term in term
| ID                            # variables and constants
| forall args, term             # dependent product
| term -> term                  # non-dependent product
| fun args => term              # lambda-abstraction
| term term                     # application
| term & term                   # intersection of types
| term | term                   # union of types
| <term,term>                   # strong pair
| proj_l term                   # left projection of a strong pair
| proj_r term                   # right projection of a strong pair
| smatch term [as ID] [return term] with ID [: term] => term, ID [: term] => term end
                                # strong sum
| inj_l term term               # left injection of a strong sum
| inj_r term term               # right injection of a strong sum
| coe term term                 # coercion
| _                             # wildcard
\end{verbatim}}
\noindent Identifiers \verb|ID| refers to any alphanumeric string (possibly with underscores and apostrophes).
The non-terminal symbol \verb|args| correspond to a non-empty sequence of arguments, where an argument is an identifier, and can be given with its type. In the latter case, you should parenthesize it, for instance {\small \verb|(x : A)|}, and if you want to assign the same type to several consecutive arguments, you can \eg write {\small \verb|(x y z : A)|}. Strong sums have a complicated syntax. For instance, consider this term:
{\small 
\begin{bull}
smatch foo as x return T with y : T1 => bar, z : T2 => baz end
\end{bull}}
\noindent The above term in the concrete syntax corresponds to
$\match{\mathtt{foo}}{\lambda x\of\_.\mathtt{T}}{y\of\mathtt{T1}}{\mathtt{bar}}{x\of\mathtt{T2}}{\mathtt{baz}}$
%
in the abstract syntax. The concrete syntax thus guarantees that the returned type is a $\lambda$-abstraction, and it allows a simplified behaviour of the type reconstruction algorithm.
The behaviour of the concrete syntax is intended to mimic Coq.


\vspace{-3mm}
\subsection{Implementation of the syntax}
\vspace{-3mm}
In the OCaml implementation, $\Delta$-terms and their types along with essences and type essences are represented with a single type called \ocamle|term|. It allows some functions (such as the normalization function) to be applied both on $\Delta$-terms and on essences.

{\small \begin{ocaml}
type term =
  | Sort of location * sort
  | Let of location * string * term * term * term (* let s : t1 := t2 in t3 *)
  | Prod of location * string * term * term (* forall s : t1, t2 *)
  | Abs of location * string * term * term (* fun s : t1 => t2 *)
  | App of location * term * term list (* t t1 t2 ... tn *)
  | Inter of location * term * term (* t1 & t2 *)
  | Union of location * term * term (* t1 | t2 *)
  | SPair of location * term * term (* < t1, t2 > *)
  | SPrLeft of location * term (* proj_l t1 *)
  | SPrRight of location * term (* proj_r t1 *)
  | SMatch of location * term * term * string * term * term * string * term * term
                    (* match t1 return t2 with s1 : t3 => t4 , s2 : t5 => t6 end *)
  | SInLeft of location * term * term (* inj_l t1 t2 *)
  | SInRight of location * term * term (* inj_r t1 t2 *)
  | Coercion of location * term * term (* coe t1 t2 *)
  | Var of location * int (* de Bruijn index *)
  | Const of location * string (* variable name *)
  | Underscore of location (* meta-variables before analysis *)
  | Meta of location * int * (term list) (* index and substitution *)
\end{ocaml}}
\noindent The constructors of \ocamle|term| contain the location information retrieved by the parser that allows the typechecker to give the precise location of a subterm to the user, in case of error.

The \ocamle|App| constructor takes as parameters the applied function and the list of all the arguments. The list of parameters is used as a stack, hence the rightmost argument is the head of the list, and can easily be removed in the OCaml recursive functions. The variables are referred to as strings in the \ocamle|Const| constructor, and as de Bruijn indices in \ocamle|Var| constructors.

The parser does not compute de Bruijn indices, it gives the variables as strings. The function \ocamle|fix_index| replaces bound variables by de Bruijn indices. We still keep track of the string names of the variables, in case we have to print them back. Its converse function, \ocamle|fix_id|, replaces the de Bruijn indices with the previous strings, possibly updating the string names in case of name clashes. For instance, the string printed to the user, showing the normalized form of \bulle|(fun (x y : nat) => x) y|, is \bulle|fun y0 : nat => y| : the bound variable \bulle|y| has been renamed \bulle|y0|. The meta-variables are generated by the typecheckers, and their identifier is an integer.

We have defined several helper functions to ease the process of terms: there is the most generic function \ocamle|visit_term f g h t|, which looks at the children of the term \ocamle|t|, and:
\begin{enumerate}
\item every child \ocamle|t1| outside of a binder is replaced with \ocamle|f t1|;
\item every child \ocamle|t1| inside the binding of a variable whose name (a string) is \ocamle|s| is replaced with \ocamle|g s t1|, while \ocamle|s| is replaced with \ocamle|h s t1|. The functions \ocamle|g| and \ocamle|h| takes a string as an argument, for helping the implementation of the \ocamle|fix_index| and \ocamle|fix_id| functions.
\end{enumerate}
The function \ocamle|map_term| is a kind of mapping function: \ocamle|map_term k f t| finds every variable \ocamle|Var(l, n)| inside the term, and replaces it by \ocamle|f (k+offset) l n|, where \ocamle|offset| is the number of extra bindings.

\begin{ocaml}
let lift k n =  map_term k (fun k l m -> if m < k then Var (l, m) else Var (l, m+n))
\end{ocaml}
%
The \ocamle|lift| and \ocamle|map_term| functions allow us to define a substitution in a clean way:
\begin{ocaml}
(* Transform (lambda x. t1) t2 into t1[t2/x] *)
let beta_redex t1 t2 =
  let subst k l m =
    if m < k then Var (l, m) (* bound variable *)
    else if m = k then (* x *)
      lift 0 k t2
    else (* the enclosing lambda goes away *)
      Var (l, m-1)
  in map_term 0 subst t1
\end{ocaml}

\vspace{-3mm}
\subsection{Environments}\label{ssec:env}
\vspace{-3mm}
There are four kinds of environments, namely:
\begin{enumerate}
\item the \emph{global environment} (noted $\Sigma$). The global environment holds constants which are fully typechecked:
 $\Sigma ::= \cdot \mid \Sigma, c \of \varsigma \loc \sigma \mid \Sigma, c := M \loc \Delta : \varsigma \loc \sigma$.
  Intuitively, $c \of \varsigma \loc \sigma$ is a declaration of a constant (or axiom), and $c := M \loc \Delta : \varsigma \loc \sigma$ corresponds to a global definition.
\item the \emph{local environment} (noted $\Gamma$). It is used for the first step of typechecking, and looks like a standard environment:
 $\Gamma ::= \cdot \mid \Gamma, x \of \sigma \mid \Gamma, x := \Delta : \sigma$.
  Intuitively, $x \of \sigma$ is a variable introduced by a $\lambda$-abstraction, and $x := \Delta : \sigma$ is a local definition introduced by a $\mathbf{let}$.
\item the \emph{essence environment} (noted $\Psi$). It is used for the second step of typechecking, and holds the essence of the local variables:
 $\Psi ::= \cdot \mid \Psi, x \mid \Psi, x := M$.
  Intuitively, $x$ is a variable introduced by a $\lambda$-abstraction, and $x := M$ is a local definition introduced by a $\mathbf{let}$. Notice that the variable $x$ in the BNF expression $\Psi, x$ carries almost no information. However, since local variables are referred to by their de Bruijn indices, and these indices are actually their position in the environment, it follows that they have to appear in the environment, even when there is no additional information.
\item the \emph{meta-environment} (noted $\Phi$). It is used for unification, and records meta-variables and their instantiation whenever the unification algorithm has found a solution:\\
    \mbox{$\Phi ::=  \cdot \mid \Phi, \mathbf{sort}(?x) \mid \Phi, ?x := s \mid \Phi, (\Gamma \vdash ?x : \s) \mid \Phi, (\Gamma \vdash ?x := \Delta : \s) \mid \Phi, \Psi \vdash ?x \mid \Phi, \Psi \vdash ?x := M$.}\\  Intuitively, since there are some meta-variables for which we know they have to be sorts, it follows that $\mathbf{sort}(?x)$ declares a meta-variable $?x$ which correspond either to $\Type$ or $\Kind$, and $?x := s$ is the instantiation of a sort $?x$. Also, $\Gamma \vdash ?x : \s$ is the declaration of a meta-variable $?x$ of type $\s$ which appeared in a local environment $\Gamma$, and $\Gamma \vdash ?x := \Delta : \s$ is the instantiation of the meta-variable $?x$. Concerning meta-variables inside essences, $\Psi \vdash ?x$ is the declaration of a meta-variable $?x$ in an essence environment $\Psi$, and $\Psi \vdash ?x := M$ is the instantiation of $?x$.
\end{enumerate}
\vspace{-3mm}
\subsection{Suspended substitution}\label{ssec:subst}
\vspace{-3mm}
We shortly introduce suspended substitution, as presented in \cite{Refine12}.
Let's consider the following example: if we want to unify $(\lambda x\of\s. ?y)\at c_1$ with $c_1$, we could unify $?y$ with $c_1$ or with $x$, the latter being the preferred solution. However, if we normalize $(\lambda x\of\s. ?y)\at c_1$, we should record the fact that $c_1$ can be substituted by any occurrence of $x$ appearing the term to be replaced by $?y$. That is the purpose of suspended substitution: the term is actually noted $(\lambda x\of\s. ?y[x])\at c_1$ and reduces to $?y[c_1]$, noting that $c_1$ has replaced $x$.

\begin{definition}[Type-erase function and suspended substitution] \hfill
  \begin{enumerate}
  \item the vector $x_1;\ldots;x_n$ is created using the type-erase function $\metastrip \cdot$, defined as\\
  $\metastrip{x_1 : \s_1;\ldots x_n : \s_n} \eqdef x_1;\ldots;x_n$.
  \item when we want to create a new meta-variable in a local context $\Gamma = x_1 : \s_1;\ldots x_n : \s_n$, we create a meta-variable $?y[\metastrip\Gamma] \equiv \mathop{?y}[x_1;\ldots;x_n]$. The vector $\D_1;\ldots;\D_n$ inside $?y[\D_1;\ldots;\D_n]$ is the suspended substitution of $?y$. Substitutions for meta-variable and their suspended substitution are propagated as follows:\\
      $\begin{array}{rcl}
        \subst{?y[\D_1;\ldots;\D_n]}{x}{\D} & \eqdef & ?y[\subst{\D_1}{x}{\D};\ldots;\subst{\D_n}{x}{\D}] \\[1mm]
        \subst{?y[M_1;\ldots;M_n]}{x}{N} & \eqdef & ?y[\subst{M_1}{x}{N};\ldots;\subst{M_n}{x}{N}]
      \end{array}$
  \end{enumerate}
\end{definition}
\vspace{-3mm}
\section{The evaluator of {\color{blue} \href{https://github.com/cstolze/Bull}{Bull}}}\label{sec:eval}
\vspace{-3mm}
The evaluator follows the applicative order strategy, which recursively normalizes all subterms from left to right (with the help of the \ocamle|visit_term| function, see full code in \cite{Bull}), then:
if the resulting term is a redex, reduces it, then use the same strategy again;
or else, the resulting term is in normal form.
\vspace{-3mm}
\subsection{Reduction rules}
\vspace{-3mm}
The reduction notions, from which we can defined one-step reduction, multistep reduction, and equivalence relation, are defined below.
\begin{enumerate}
\item for $\D$-terms:\\
    $\begin{array}{rll}
      (\lambda x\of\s.\Delta_1)\at\D_2 & \notion_\beta & \subst{\Delta_1}{x}{\Delta_2} \\[1mm]
      \lambda x\of\s.\Delta\at x & \notion_\eta & \Delta \hfill\IF x\not\in\FV(\D)\\[1mm]
      \pri \spair{\D_1}{\D_2} & \notion_{\pri} & \D_i \\[1mm]
      \multicolumn{3}{l}{\match{\inni \D_3}{\s}{x\of\t}{\D_1}{x\of\r}{\D_2}}\\[1mm] 
        & \notion_{\inni} & \subst{\D_i}{x}{\D_3}\\[3mm]
      \llet{x}{\s}{\D_1}{\D_2} & \notion_\zeta & \subst{\D_2}{x}{\D_1} \\[1mm]
      c & \notion_{\delta\Sigma} & \D \hfill\IF (c := M \loc \Delta : \varsigma \loc \sigma) \in \Sigma  \\[1mm]
      x & \notion_{\delta\Gamma} & \D \hfill\IF (x := \Delta : \sigma) \in \Gamma  \\[1mm]
      ?x[\D_1;\ldots;\D_n] & \notion_{\delta\Phi} & \vecsubst{\D}{\D_i}{\metastrip\Gamma} \phantom{\hspace{1cm}}\hfill\IF (\Gamma \vdash ?x := \Delta : \s) \in \Phi  \\[1mm]
      ?x[\D_1;\ldots;\D_n] & \notion_{\delta\Phi} & s \hfill\IF (\Gamma \vdash ?x := s) \in \Phi  
    \end{array}
    $

\item for pure $\lambda$-terms:\\
    $\begin{array}{@{\qquad}rcl}
      (\lambda x.M)\at N & \notion_\beta & \subst{M}{x}{N} \\[1mm]
      \lambda x.M\at x & \notion_\eta & M \hfill\IF x\not\in\FV(M) \\[1mm]
      \llett{x}{M}{N} & \notion_\zeta & \subst{N}{x}{M} \\[1mm]
      c & \notion_{\delta\Sigma} & M  \phantom{\hspace{1.4cm}}\hfill\IF (c := M \loc \Delta : \varsigma \loc \sigma) \in \Sigma  \\[1mm]
      x & \notion_{\delta\Psi} & M \hfill\IF (x := M) \in \Psi  \\[1mm]
      ?x[M_1;\ldots;M_n] & \notion_{\delta\Phi} & \vecsubst{N}{M_i}{\metastrip\Psi} \hfill\IF (\Psi \vdash ?x := M) \in \Phi  \\[3mm]
    \end{array}$
\end{enumerate}
\vspace{-3mm}
\subsection{Implementation}
\vspace{-3mm}
When the user inputs a term, the refiner creates meta-variables and tries to instantiate them, but this should remain as much as possible invisible to the user. Therefore the term returned by the refiner should be meta-variable free, even though not in normal form. Thus terms in the global signature $\Sigma$ are meta-variable free, and the $\delta\Phi$ reductions are only used by the unifier and the refiner.

If we want to normalize a term, The function \ocamle|strongly_normalize| works on both $\D$-terms and pure $\lambda$-terms, and supposes that the given term is meta-variable free. Note that reductions can create odd spines, for instance if you consider the term $(\lambda x\of\s. x\at S_1)\at(\D\at S_2)$, a simple $\beta$-redex would give $\D\at S_2\at S_1$, therefore we merge $S_2$ and $S_1$ in a single spine.
\begin{ocaml}
let rec strongly_normalize is_essence env ctx t =
  let sn_children = visit_term (strongly_normalize is_essence env ctx)
			       (fun _ -> strongly_normalize is_essence
                         $\,\,$env (Env.add_var ctx (DefAxiom ("",nothing))))
			       (fun id _ -> id)
  in let sn = strongly_normalize is_essence env ctx in
  (* Normalize the children *)
  let t = sn_children t in
  match t with
  (* Spine fix *)
  | App(l, App(l',t1,t2), t3) ->
     sn (App(l, t1, List.append t2 t3))
  (* Beta-redex *)
  | App (l, Abs (l',_,_, t1), t2 :: [$\,$]) ->
     sn (beta_redex t1 t2)
  | App (l, Abs (l',x,y, t1), t2 :: t3)
    -> sn @@ app l (sn (App(l,Abs (l',x,y, t1), t3))) t2
  | Let (l, _, t1, t2, t3) -> sn (beta_redex t2 t1)
  (* Delta-redex *)
  | Var (l, n) -> let (t1, _) = Env.find_var ctx n in
	     (match t1 with
	      | Var _ -> t1
	      | _ -> sn t1)
  | Const (l, id) -> let o = Env.find_const is_essence env id in
	             (match o with
                 | None -> Const(l, id)
	              | Some (Const (_,id') as t1,_) when id = id' -> t1
	              | Some (t1,_) -> sn t1)
  (* Eta-redex *)
  | Abs (l,_, _, App (l',t1, Var (_,0) :: l2))
    -> if is_eta (App (l', t1, l2)) then
         let t1 = lift 0 (-1) t1 in
         match l2 with
         | [$\,$] -> t1
         | _ -> App (l', t1, List.map (lift 0 (-1)) l2)
       else t
  (* Pair-redex *)
  | SPrLeft (l, SPair (l', x,_)) -> x
  | SPrRight (l, SPair (l', _, x)) -> x
  (* inj-reduction *)
  | SMatch (l, SInLeft(l',_,t1), _, id1, _, t2, id2, _, _) ->
     sn (beta_redex t2 t1)
  | SMatch (l, SInRight(l',_,t1), _, id1, _, _, id2, _, t2) ->
     sn (beta_redex t2 t1)
  | _ -> t
\end{ocaml}

\vspace{-3mm}
\section{The subtyping algorithm of {\color{blue} \href{https://github.com/cstolze/Bull}{Bull}}}\label{sec:sub}
\vspace{-3mm}
The subtyping algorithm $\cal A$  is basically the same as the one described and Coq certified/extracted in \cite{TTCS17}. The only difference is that the types are normalized before applying the algorithm. The auxiliary  rewriting functions ${\cal R}_1,{\cal R}_2,{\cal R}_3,{\cal R}_4$, described in \cite{TTCS17}, rewrite terms in normal forms as follows:
\begin{ocaml}
let rec anf a =
  let rec distr f a b =
    match (a,b) with
    | (Union(l,a1,a2),_) -> Inter(l, distr f a1 b, distr f a2 b)
    | (_, Inter(l,b1,b2)) -> Inter(l, distr f a b1, distr f a b2)
    | _ -> f a b
  in
  match a with
  | Prod(l,id,a,b) -> distr (fun a b -> Prod(l,id,a,b)) (danf a) (canf b)
  | _ -> a
and canf a =
  let rec distr a b =
    match (a,b) with
    | (Inter(l,a1,a2),_) -> Inter(l, distr a1 b, distr a2 b)
    | (_,Inter(l,b1,b2)) -> Inter(l, distr a b1, distr a b2)
    | _ -> Union(dummy_loc,a,b)
  in
  match a with
  | Inter(l,a,b) -> Inter(l, canf a, canf b)
  | Union(l,a,b) -> distr (canf a) (canf b)
  | _ -> anf a
and danf a =
  let rec distr a b =
    match (a,b) with
    | (Union(l,a1,a2),_) -> Union(l, distr a1 b, distr a2 b)
    | (_,Union(l,b1,b2)) -> Union(l, distr a b1, distr a b2)
    | _ -> Inter(dummy_loc, a,b)
  in
  match a with
  | Inter(l,a,b) -> distr (danf a) (danf b)
  | Union(l,a,b) -> Union(l, danf a, danf b)
  | _ -> anf a
\end{ocaml}
It follows that, our subtyping function is quite simple:
\begin{ocaml}
let is_subtype env ctx a b =
  let a = danf @@ strongly_normalize false env ctx a in
  let b = canf @@ strongly_normalize false env ctx b in
  let rec foo env ctx a b =
  match (a, b) with
  | (Union(_,a1,a2),_) -> foo env ctx a1 b && foo env ctx a2 b
  | (_,Inter(_,b1,b2)) -> foo env ctx a b1 && foo env ctx a b2
  | (Inter(_,a1,a2),_) -> foo env ctx a1 b || foo env ctx a2 b
  | (_,Union(_,b1,b2)) -> foo env ctx a b1 || foo env ctx a b2
  | (Prod(_,_,a1,a2),Prod(_,_,b1,b2))
    -> foo env ctx b1 a1 && foo env (Env.add_var ctx (DefAxiom("",nothing))) a2 b2
  | _ -> same_term a b
  in foo env ctx a b
\end{ocaml}

\vspace{-3mm}
\section{The unification algorithm of {\color{blue} \href{https://github.com/cstolze/Bull}{Bull}}}\label{sec:unif}
\vspace{-3mm}
Higher-order unification of two terms $\D_1$ and $\D_2$ aims at finding a most general substitution for meta-variables such that $\D_1$ and $\D_2$ becomes convertible. The structural rules are given in Figure \ref{struct-unif}. Classical references are the work of Huet \cite{huet75}, and Dowek, Kirchner, and Hardin\,\cite{dowekunif}.

Our higher-order unification algorithm is inspired by the Reed \cite{reed-unif} and Ziliani-Sozeau \cite{ziliani-unif} papers. In \cite{ziliani-unif}, conversion of terms is quite involved because of the complexity of Coq. For simplicity, our algorithm supposes the terms to be in normal form.

The unification algorithm takes as input a meta-environment $\Phi$, a global environment $\Sigma$, a local environment $\Gamma$, the two terms to unify $\D_1$ and $\D_2$, and either fails or returns the updated meta-environment $\Phi$. 
The rest of the unification algorithm implements \emph{Higher-Order Pattern Unification} (HOPU) \cite{reed-unif}. In a nutshell, HOPU takes as an argument a unification problem $?f\at S \unif N$, where all the terms in $S$ are free variables and each variable occurs once.
For instance, for the unification problem $?f\at y\at x\at z \unif x\at c\at y$, it creates the solution $?f := \lambda y\of\s_2.\lambda x\of \s_1.\lambda z\of\s_3.  x\at c\at y$. The expected type of $x$, $y$, and $z$ can be found in the local environment, but capturing correctly the free variables $x$, $y$, and $z$ is quite tricky because we have to permute their de Bruijn indices.
If HOPU does not work, we try to recursively unify every subterm.

\begin{figure}[t!]
  \hspace{-4cm}
    $\begin{array}{c}
      \infer[(\mathit{Sort})]{\Phi; \Sigma; \Gamma \vdash s_1 \unif s_2 \unifresult \Phi}{s_1 \equiv s_2}
      \quad
      \infer[(\mathit{Const})]{\Phi; \Sigma; \Gamma \vdash c_1 \unif c_2 \unifresult \Phi}{c_1 \equiv c_2}
      \quad
      \infer[(\mathit{Var})]{\Phi; \Sigma; \Gamma \vdash x_1 \unif x_2 \unifresult \Phi}{x_1 \equiv x_2}
      \\[1mm]
      \infer[(\mathit{Abs})]{\Phi_1; \Sigma; \Gamma \vdash \lambda x\of\s_1.\Delta_1 \unif \lambda x\of\s_2. \Delta_2 \unifresult \Phi_3}{
        \Phi_1; \Sigma; \Gamma \vdash \s_1 \unif \s_2 \unifresult \Phi_2
        & \Phi_2; \Sigma; \Gamma, x\of\s_1 \vdash \Delta_1 \unif \Delta_2 \unifresult \Phi_3
      }
      \\[1mm]
      \infer[(\cap)]{\Phi_1; \Sigma; \Gamma \vdash \s_1 \cap \t_1 \unif \s_2 \cap \t_2 \unifresult \Phi_3}{
        \Phi_1; \Sigma; \Gamma \vdash \s_1 \unif \s_2 \unifresult \Phi_2
        & \Phi_2; \Sigma; \Gamma \vdash \t_1 \unif \t_2 \unifresult \Phi_3
      }
      \qquad 
      \infer[(\cup)]{\Phi_1; \Sigma; \Gamma \vdash \s_1 \cup \t_1 \unif \s_2 \cup \t_2 \unifresult \Phi_3}{
        \Phi_1; \Sigma; \Gamma \vdash \s_1 \unif \s_2 \unifresult \Phi_2
        & \Phi_2; \Sigma; \Gamma \vdash \t_1 \unif \t_2 \unifresult \Phi_3
      }
      \\[1mm]
      \infer[(\mathit{Spair})]{\Phi_1; \Sigma; \Gamma \vdash \spair{\Delta_1}{\Delta_3} \unif \spair{\Delta_2}{\Delta_4} \unifresult \Phi_3}{
        \Phi_1; \Sigma; \Gamma \vdash \Delta_1 \unif \Delta_2 \unifresult \Phi_2
        & \Phi_2; \Sigma; \Gamma \vdash \Delta_3 \unif \Delta_4 \unifresult \Phi_3
      }
      \qquad 
      \infer[(\mathit{Proj})]{\Phi_1; \Sigma; \Gamma \vdash \pri \Delta_1 \unif \pri \Delta_2 \unifresult \Phi_2}{
        \Phi_1; \Sigma; \Gamma \vdash \s_1 \unif \s_2 \unifresult \Phi_2
      }
      \\[1mm]
      \infer[(\mathit{Inj})]{\Phi_1; \Sigma; \Gamma \vdash \inni \s_1 \Delta_1 \unif \inni \s_2 \Delta_2 \unifresult \Phi_3}{
        \Phi_1; \Sigma; \Gamma \vdash \s_1 \unif \s_2 \unifresult \Phi_2
        & \Phi_2; \Sigma; \Gamma \vdash \Delta_1 \unif \Delta_2 \unifresult \Phi_3
      }
      \qquad 
      \infer[(\mathit{Coe})]{\Phi_1; \Sigma; \Gamma \vdash \coe\at \s_1\at \Delta_1 \unif \coe\at \s_2\at \Delta_2 \unifresult \Phi_3}{
        \Phi_1; \Sigma; \Gamma \vdash \s_1 \unif \s_2 \unifresult \Phi_2
        & \Phi_2; \Sigma; \Gamma \vdash \Delta_1 \unif \Delta_2 \unifresult \Phi_3
      }
      \\[3mm] 
      \infer[(\mathit{Ssum})]{
        \begin{array}{r}
          \Phi_1; \Sigma; \Gamma \vdash \match{\Delta}{\tau}{x\of\s_1}{\Delta_1}{x\of\s_2}{\Delta_2} 
          \unif \match{\Delta'}{\tau'}{x\of\s'_1}{\Delta_1}{x\of\s'_2}{\Delta'_2} \unifresult \Phi_7
      \end{array}}{
        \begin{array}{l@{\qquad}l@{\qquad}l}
          \Phi_1; \Sigma; \Gamma \vdash \Delta \unif \Delta' \unifresult \Phi_2 &
          \Phi_1; \Sigma; \Gamma \vdash \tau \unif \tau' \unifresult \Phi_3 &
          \Phi_1; \Sigma; \Gamma \vdash \s_1 \unif \s'_1 \unifresult \Phi_4 \\[0mm]
          \Phi_1; \Sigma; \Gamma, x\of\s_1 \vdash \Delta_1 \unif \Delta'_1 \unifresult \Phi_5 &
          \Phi_1; \Sigma; \Gamma \vdash \s_2 \unif \s'_2 \unifresult \Phi_6 &
          \Phi_1; \Sigma; \Gamma, x\of\s_2 \vdash \Delta_2 \unif \Delta'_2 \unifresult \Phi_7
        \end{array}
      }
      \\[2mm] 
      \infer[(\eta_l)]{\Phi_1; \Sigma; \Gamma \vdash \lambda x\of\s. \Delta_1 \unif \Delta_2 \unifresult \Phi_2}{
        \Phi_1; \Sigma; \Gamma, x\of\s \vdash \Delta_1 \unif \Delta_2\, x \unifresult \Phi_2
        & \mbox{$\Delta_2$ is not a $\lambda$-abstraction}
      }
      \quad 
      \infer[(\eta_r)]{\Phi_1; \Sigma; \Gamma \vdash \Delta_1 \unif \lambda x\of\s.\Delta_2 \unifresult \Phi_2}{
        \Phi_1; \Sigma; \Gamma, x\of\s \vdash \Delta_1\, x \unif \Delta_2 \unifresult \Phi_2
        & \mbox{$\Delta_1$ is not a $\lambda$-abstraction}
      }
    \end{array}
    $
  \vspace{-3mm}
  \caption{Structural rules of the unification algorithm}\label{struct-unif}
   \vspace{-3mm}
\end{figure}

\vspace{-3mm}
\section{The refinement algorithm of {\color{blue} \href{https://github.com/cstolze/Bull}{Bull}}}\label{sec:refine}
\vspace{-3mm}
The {\color{blue} \href{https://github.com/cstolze/Bull}{Bull}} refinement algorithm is inspired by the work on the Matita ITP \cite{Refine12}. It is defined using \emph{bi-directionality}, in the style of Harper-Licata \cite{harper-licata}. The bi-directional technique is a mix of typechecking and type reconstruction, in order to trigger the unification algorithm as soon as possible. Moreover, it gives more precise error messages than standard type reconstruction. For instance, if \bulle|f : (bool -> nat -> bool) -> bool|, then \bulle|f (fun x y => y)| is ill-typed. With a simple type inference algorithm, we would type \bulle|f|, then \bulle|fun x y => y| which would be given some type \bulle|?x -> ?y -> ?y|, and finally we would try to unify \bulle|bool -> nat -> bool| with \bulle|?x -> ?y -> ?y|, which fails. However, the failure is localized on the application, whereas it would better be localized inside the argument. More precisely, we would have the following error message:
\begin{bull}
f (fun x y => y)
              ^
Error: the term "y" has type "nat" while it is expected to have type "bool".
\end{bull}
Our typechecker is also a \emph{refiner}: intuitively, a refiner takes as input an incomplete term, and possibly an incomplete type, and tries to infer as much information as possible in order to reconstruct a well-typed term.
For example, let's assume we have in the global environment the following constants:
\begin{bull}
(eq : nat -> nat -> Type), (eq_refl : forall x : nat, eq x x)
\end{bull}
Then refining the term {\small \verb|eq_refl _ : eq _ 0|} would create the following term:
\begin{bull}
eq_refl 0 : eq 0 0
\end{bull}
Refinement also enable untyped abstractions: the refiner may recover the type of bound variables, because untyped abstractions are incomplete terms.
\noindent The typechecking is done in two steps: firstly the term is typechecked without caring about the essence, then we check the essence.
The five typing judgment are defined as follows:

\begin{figure}[t!]
\hspace{-1cm}
  $\begin{array}{c}
    \infer[(T)]{\Phi; \Sigma; \Gamma \vdash \Type \uprefine \Type : \Kind; \Phi}{}
    \\[1mm]
    \infer[(\mathit{Let})]{\Phi_1; \Sigma; \Gamma \vdash \llet{x}{\s}{\Delta_1}{\Delta_2} \uprefine \llet{x}{\s'}{\Delta'_1}{\Delta'_2} : \subst{\tau}{x}{\s'}; \Phi_4}{
        \Phi_1; \Sigma; \Gamma \vdash \sigma \forcetype \s'; \Phi_2 &
        & 
        \Phi_2; \Sigma; \Gamma \vdash \Delta : \s' \downrefine \Delta'; \Phi_3
        &
        \Phi_3; \Sigma; \Gamma, x := \Delta'\of\sigma' \vdash \Delta_2 \uprefine \Delta'_2 : \tau; \Phi_4
    }
    \\[1mm]
    \infer[(\mathit{Prod})]{\Phi_1; \Sigma; \Gamma \vdash \Pi x\of\s_1.\s_2 \uprefine \Pi x\of\s'_1.\s'_2 : s_2; \Phi_4}{
    \Phi_1; \Sigma; \Gamma \vdash \s_1 \forcetype \s'_1 : s_1; \Phi_2
    &
      \Phi_2; \Sigma; \Gamma \vdash \s_2 \forcetype \s'_2 : s_2; \Phi_3
    &
      \Phi_3 \vdash (s_1, s_2) \in \PTS; \Phi_4
      }
    \\[1mm]
      \infer[(\mathit{Abs})]{\Phi_1; \Sigma; \Gamma \vdash \lambda x\of\s.\Delta \uprefine \lambda x\of\s'.\Delta' : \Pi x\of\s'.\t; \Phi_4}{
        \Phi_1; \Sigma; \Gamma \vdash \s \forcetype \s'; \Phi_2 &
        & 
        \Phi_2; \Sigma; \Gamma, x\of\s' \vdash \Delta \uprefine \Delta' : \t; \Phi_3
        & \Phi_3; \Sigma; \Gamma \vdash \Pi x\of\s'.\t \forcetype \r : s; \Phi_4
      }
    \\[1mm]
    \infer[(App_1)]{\Phi_1; \Sigma; \Gamma \vdash \Delta\at () \uprefine \Delta' : \s ; \Phi_2}{
      \Phi_1; \Sigma; \Gamma \vdash \Delta \uprefine \Delta' : \s; \Phi_2
    }
    \\[1mm]
    \infer[(App_2)]{\Phi_1; \Sigma; \Gamma \vdash \Delta_1\at (S; \Delta_2) \uprefine \Delta'_1\at\Delta'_2 : \subst{\s_2}{x}{\Delta'_2} ; \Phi_3}{
      \Phi_1; \Sigma; \Gamma \vdash \Delta_1\at S \uprefine \Delta' : \s; \Phi_2 
      & 
      \Phi_2; \Sigma; \Gamma \vdash \s =_\beta \Pi x\of\s_1. \s_2 &
      \Phi_2; \Sigma; \Gamma \vdash \Delta_2 : \s_1 \downrefine \Delta'_2; \Phi_3
    }
    \\[1mm]
    \infer[(App_3)]{\Phi_1; \Sigma; \Gamma \vdash \Delta_1\at (S; \Delta_2) \uprefine \Delta'\at\Delta'_2 : \subst{?x[\metastrip{\Gamma};x]}{x}{\Delta'_2} ; \Phi_4}{
      \begin{array}{ll}
      \Phi_1; \Sigma; \Gamma \vdash \Delta_2\at S \uprefine \Delta' : \s; \Phi_2 &
      \Phi_2; \Sigma; \Gamma \vdash \Delta_2 \uprefine \Delta'_2 : \s_1; \Phi_3 \\[0mm]
      \multicolumn{2}{c}{
      \Phi_3, ?y, (\Gamma, x\of\s_1 \vdash ?x : ?y[\,]); \Sigma; \Gamma \vdash \s \unif \Pi x\of\s_1. ?x[\metastrip{\Gamma}; x] \unifresult \Phi_4}
      \end{array}
    }
  \end{array}
  $
    \vspace{-3mm}
\caption{Rules for $\protect\uprefine$ (1st part)}\label{uprefine1}
  \vspace{-3mm}
\end{figure}

\begin{figure}[t!]
\hspace{-1cm}
  $\begin{array}{c}
      \infer[(\cap)]{\Phi_1; \Sigma; \Gamma \vdash \s_1 \cap \s_2 \uprefine \s'_1 \cap \s'_2 : \Type; \Phi_3}
      { \Phi_1; \Sigma; \Gamma \vdash \s_1 : \Type \downrefine \s'_1; \Phi_2
    & \Phi_2; \Sigma; \Gamma \vdash \s_2 : \Type \downrefine \s'_2; \Phi_3}
    \\[1mm]
    \infer[(\cup)]{\Phi_1; \Sigma; \Gamma \vdash \s_1 \cup \s_2 \uprefine \s'_1 \cup \s'_2 : \Type; \Phi_3}
      { \Phi_1; \Sigma; \Gamma \vdash \s_1 : \Type \downrefine \s'_1; \Phi_2
      & \Phi_2; \Sigma; \Gamma \vdash \s_2 : \Type \downrefine \s'_2; \Phi_3}
    \\[1mm]
    \infer[(\mathit{Spair})]{\Phi_1; \Sigma; \Gamma \vdash \spair{\Delta_1}{\Delta_2} \uprefine \spair{\Delta'_1}{\Delta'_2} : \s_1 \cap \s_2; \Phi_4}{
    \Phi_1; \Sigma; \Gamma \vdash \Delta_1 \uprefine \Delta'_1 : \s_1; \Phi_2 
    & 
    \Phi_2; \Sigma; \Gamma \vdash \Delta_2 \uprefine \Delta'_2 : \s_2; \Phi_3 &
    \Phi_3; \Sigma; \Gamma \vdash \sigma_1 \cap \sigma_2 : \Type \downrefine \Phi_4
    }
    \\[1mm]
    \infer[(Proj_1)]{\Phi_1; \Sigma; \Gamma \vdash \pri \Delta \uprefine \pri \Delta' : \s_i; \Phi_2}{
    \Phi_1; \Sigma; \Gamma \vdash \Delta \uprefine \Delta' : \s; \Phi_2
    & \Phi_1; \Sigma; \Gamma \vdash \s =_\beta \s_1 \cap \s_2
      }
    \\[1mm]
    \infer[(Proj_2)]{\Phi_1; \Sigma; \Gamma \vdash \pri \Delta \uprefine \pri \Delta' : ?x_i[\metastrip{\Gamma}]; \Phi_3}{
    \Phi_1; \Sigma; \Gamma \vdash \Delta \uprefine \Delta' : \s; \Phi_2
    & 
    \Phi_2, (\Gamma \vdash ?x_1 : \Type), (\Gamma \vdash ?x_2 : \Type); \Sigma; \Gamma \vdash \s \unif ?x_1[\metastrip{\Gamma}] \cap ?x_2[\metastrip{\Gamma}] \unifresult \Phi_3
    }
    \\[3mm] 
    \infer[(\mathit{Ssum})]{
      \begin{array}{l}
        \Phi_1; \Sigma; \Gamma \vdash \match{\Delta}{\lambda x\of\tau_1.\tau_2}{x\of\s_1}{\Delta_1}{x\of\s_2}{\Delta_2} \uprefine \\[0mm]
        \quad \match{\Delta'}{\t'}{x\of\s'_1}{\Delta'_1}{x\of\s'_2}{\Delta'_2} : \subst{\t'_2}{x}{\Delta'}; \Phi_8
    \end{array}}{
      \begin{array}{ll}
      \Phi_1; \Sigma; \Gamma \vdash \Delta \uprefine \Delta' : \s'; \Phi_2 &
      \Phi_2; \Sigma; \Gamma \vdash \lambda x\of\tau_1.\tau_2 : \Pi x\of\s \rightarrow \Type \downrefine \lambda x\of\tau'_1.\tau'_2; \Phi_3 \\[0mm]
      \Phi_3; \Sigma; \Gamma \vdash \s_1 : \Type \downrefine \s'_1; \Phi_4 &
      \Phi_4; \Sigma; \Gamma \vdash \s_2 : \Type \downrefine \s'_2; \Phi_5 \\[0mm]
      \Phi_5; \Sigma; \Gamma \vdash \s' \unif \s'_1 \cup \s'_2 \unifresult \Phi_6 &
      \Phi_6; \Sigma; \Gamma, x\of\s'_1 \vdash \Delta_1 : \subst{\t'_2}{x}{\innl\at\s'_2\at x} \downrefine \Delta'_1; \Phi_7 \\[0mm]
      \multicolumn{2}{l}{
      \Phi_7; \Sigma; \Gamma, x\of\s'_2 \vdash \Delta_2 : \subst{\t'_2}{x}{\innr\at\s'_1\at x} \downrefine \Delta'_2; \Phi_8}
      \end{array}
    }
    \\[4mm] 
    \infer[(\mathit{Coe})]{\Phi_1; \Sigma; \Gamma \vdash \coe\at \s\at \Delta \uprefine \coe\at \s'\at \Delta' : \s'; \Phi_3}{
      \Phi_1; \Sigma; \Gamma \vdash \s \forcetype \s' : s; \Phi_2 &
      \Phi_2; \Sigma; \Gamma \vdash \Delta \uprefine \Delta' : \t; \Phi_3 &
      \Sigma; \Gamma \vdash \t \leq \s'
    }
    \\[1mm]
    \infer[(\mathit{Var})]{\Phi; \Sigma; \Gamma \vdash x \uprefine x : \s; \Phi}{
    (x\of\s) \in \Gamma \mbox{ or } (x := \Delta : \s) \in \Gamma
    }
    \qquad \qquad 
    \infer[(\mathit{Const})]{\Phi; \Sigma; \Gamma \vdash c \uprefine c : \s; \Phi}{
    (c\of\s) \in \Sigma \mbox{ or } (c := \Delta : \s) \in \Sigma
    }
    \\[2mm]
    \infer[(\mathit{Wildcard})]{\Phi; \Sigma; \Gamma \vdash \_ \uprefine ?x[\metastrip{\Gamma}] : ?y[\metastrip{\Gamma}]; \Phi, ?z, (\Gamma \vdash ?y : ?z[\,]), (\Gamma \vdash ?x : ?y[\metastrip{\Gamma}])}{}
    \\[2mm]
    \infer[(\mathit{Meta{-}Var})]{\Phi; \Sigma; \Gamma \vdash ?x[\Delta_1;\ldots;\Delta_n] \uprefine ?x[\Delta_1;\ldots;\Delta_n] : \vecsubst{\s}{\metastrip{\Gamma'}}{\Delta_i}; \Phi}{
        (\Gamma' \vdash ?x : \s) \in \Phi \OR (\Gamma' \vdash ?x := \Delta : \s) \in \Phi
        & \Gamma' = \s_1,\ldots,\s_n
        & \Phi; \Sigma; \Gamma \vdash \D_i : \s_i \quad (i = 1\ldots n)
    }
  \end{array}
  $
    \vspace{-3mm}
\caption{Rules for $\protect\uprefine$ (2nd part)} \label{uprefine2}
  \vspace{-3mm}
\end{figure}
\begin{definition}[Typing judgments]
  We have five typing judgments, corresponding to five OCaml functions:
  \begin{enumerate}
  \item The function \ocamle|reconstruct| takes as inputs a meta-environment $\Phi_1$, a global environment $\Sigma$, a local environment $\Gamma$, and a term $\D_1$. It either fails or fills the holes in $\D_1$, which becomes $\D_2$, and returns $\D_2$ along with its type $\s$ and the updated meta-environment $\Phi_2$. The corresponding judgment is the following 
 $\Phi_1; \Sigma; \Gamma \vdash \Delta_1 \uprefine \Delta_2 : \sigma; \Phi_2$, 
and the rules are described in Figures \ref{uprefine1} and \ref{uprefine2};

  \item The function \ocamle|force_type| takes as inputs a meta-environment $\Phi_1$, a global environment $\Sigma$, a local environment $\Gamma$, and a term $\s_1$. It either fails or fills the holes in $\s_1$, which becomes $\s_2$ while ensuring it is a type, \ie its type is a sort $s$, and returns $\s_2$ along with $s$, and the updated meta-environment $\Phi_2$. The corresponding judgment is the following
 $\Phi_1; \Sigma; \Gamma \vdash \s_1 \forcetype \s_2 : \t; \Phi_2$,
and the rules are described in Figure \ref{forcetype}. Intuitively, the function reconstruct the type $\t$ of $\s_1$, then tries to unify $\t$ with $\Type$ and $\Kind$. If it can only do one unification, it keeps the successful one, if both unifications work, we choose unification with a sort meta-variable, so $\t$ is convertible to a sort;

\item The function \ocamle|reconstruct_with_type| takes as inputs a meta-environment $\Phi_1$, a global environment $\Sigma$, a local environment $\Gamma$, a term $\D_1$, and its expected type $\s$. It either fails or fills the holes in $\D_1$, which becomes $\D_2$ while ensuring its type is $\s$, and returns $\D_2$ along the updated meta-environment $\Phi_2$. The corresponding judgment is the following
 $\Phi_1; \Sigma; \Gamma \vdash \Delta_1 : \sigma \downrefine \Delta_2; \Phi_2$,
and the rules are described in Figure \ref{downrefine}. There is a rule $(\mathit{Default})$ which applies only if none of the other rules work. The acute reader could remark  two subtle things:
    \begin{enumerate}
    \item we chose not to add any inference rule for coercions, because we believe it would make error messages clearer: more precisely, if we want to check that $\coe\at\s\at\D$ has type $\t$, there could be two errors happening concurrently: it is possible that the type of $\D$ is not a subtype of $\s$, and at the same time $\s$ is not unifiable with $\t$. We think that the error to be reported should be the first one, and in this case the $(\mathit{Default})$ rule is sufficient;
      \item the management of de Bruijn indices for the $(\mathit{Let})$ is tricky: if we want to check that $\llet{x}{\s}{\Delta_1}{\Delta_2}$ has type $\t$ in some local context $\Gamma$, we recursively check that $\Delta_2$ has type $\t$ in the local context $\Gamma, x := \Delta'_1 : \s'$ for some $\D'_1$, but the de Bruijn indices for $\t$ correspond to the position of the local variables in the local context, which has been updated. We therefore have to increment all the de Bruijn indices in $\t$, in order to report the fact that there is one extra element in the local context; 
    \end{enumerate}

  \item The function \ocamle|essence| takes as inputs a meta-environment $\Phi_1$, a global environment $\Sigma$, an essence environment $\Psi$, and a term $\D$. It either fails or construct its essence $M$, and returns $M$ along with the updated meta-environment $\Phi_2$. The corresponding judgment is the following
 $\Phi_1; \Sigma; \Psi \vdash \Delta \upessence M; \Phi_2$,
and the rules are described in Figure \ref{upessence};

  \item The function \ocamle|essence_with_hint| takes as inputs a meta-environment $\Phi_1$, a global environment $\Sigma$, an essence environment $\Psi$, a term $\D$, and its expected essence $M$. It either fails or succeeds by returning the updated meta-environment $\Phi_2$. The corresponding judgment is the following
$\Phi_1; \Sigma; \Psi \vdash M \loc \Delta \downessence \Phi_2$,
and the rules are described in Figure \ref{downessence}. There is a rule $(\mathit{Default})$ which applies only if none of the other rules work.
  \end{enumerate}
\end{definition}


\begin{figure}[t!]
\hspace{-2.5cm}
  $\begin{array}{c}
    \infer[(\mathit{Force_1})]{\Phi_1; \Sigma; \Gamma \vdash \s \forcetype \s' : \t; \Phi_4}{
      \begin{array}{llll}
      \Phi_1; \Sigma; \Gamma \vdash \s \uprefine \s' : \t; \Phi_2
      &
      \Phi_2; \Sigma; \Gamma \vdash \t \unif \Type \unifresult \Phi_3
      & 
      \Phi_2; \Sigma; \Gamma \vdash \t \unif \Kind \unifresult \Phi'_3
      &
      \Phi_2, \mathbf{sort}(?x); \Sigma \vdash \t\unif s \unifresult \Phi_4
      \end{array}
      }
    \\[1mm]
    \infer[(\mathit{Force_2})]{\Phi_1; \Sigma; \Gamma \vdash \s \forcetype \s' : \t; \Phi_3}{
    \Phi_1; \Sigma; \Gamma \vdash \s \uprefine \s' : \t; \Phi_2
    &
      \Phi_2; \Sigma; \Gamma \vdash \t \unif \Type \unifresult \Phi_3
    &
      \Phi_2; \Sigma; \Gamma \notvdash \t \unif \Kind \unifresult \Phi'_3
      }
    \\[1mm]
    \infer[(\mathit{Force_3})]{\Phi_1; \Sigma; \Gamma \vdash \s \forcetype \s' : \t; \Phi'_3}{
    \Phi_1; \Sigma; \Gamma \vdash \s \uprefine \s' : \t; \Phi_2
    &
      \Phi_2; \Sigma; \Gamma \notvdash \t \unif \Type \unifresult \Phi_3
    &
      \Phi_2; \Sigma; \Gamma \vdash \t \unif \Kind \unifresult \Phi'_3
      }
  \end{array}
  $
  \vspace{-3mm}
\caption{Rules for $\protect\forcetype$} \label{forcetype}
  \vspace{-3mm}
\end{figure}

\begin{figure}[t!]
\hspace{-2.5cm}
  $\begin{array}{c}
    \infer[(\mathit{Default})]{\Phi_1; \Sigma; \Gamma \vdash \Delta : \t \downrefine \Delta'; \Phi_3}{
      \Phi_1; \Sigma; \Gamma \vdash \Delta \uprefine \Delta' : \s; \Phi_2
      &
      \Phi_2; \Sigma; \Gamma \vdash \s \unif \t \unifresult \Phi_3
    }
    \\[1mm]
    \infer[(\mathit{Let})]{\Phi_1; \Sigma; \Gamma \vdash \llet{x}{\s}{\Delta_1}{\Delta_2} : \t \downrefine \llet{x}{\s}{\Delta_1}{\Delta_2}; \Phi_4}{
        \Phi_1; \Sigma; \Gamma \vdash \s \forcetype \s' : s; \Phi_2
        & 
        \Phi_2; \Sigma; \Gamma \vdash \Delta_1 : \s' \downrefine \Delta'_1; \Phi_3
        &
        \Phi_3; \Sigma; \Gamma, x := \Delta'_1 : \s' \vdash \Delta_2 : \t \downrefine \Delta'_2; \Phi_4
    }
    \\[1mm]
    \infer[(\mathit{Abs})]{\Phi_1; \Sigma; \Gamma \vdash \lambda x : \s. \Delta : \t \downrefine \lambda x : \s'. \Delta'; \Phi_4}{
      \begin{array}{llll}
        \Phi_1; \Sigma; \Gamma \vdash \t =_\beta \Pi x : \t_1. \t_2
        & \Phi_1; \Sigma; \Gamma \vdash \s \forcetype \s'; \Phi_2 
        & \Phi_2; \Sigma; \Gamma \vdash \s' \unif \t_1 \unifresult \Phi_3
        & \Phi_3; \Sigma; \Gamma, x\of\s' \vdash \Delta : \t_2 \downrefine \Delta'; \Phi_4
      \end{array}
    }
    \\[1mm]
    \infer[(\mathit{Spair})]{\Phi_1; \Sigma; \Gamma \vdash \spair{\Delta_1}{\Delta_2} : \s \downrefine \spair{\Delta'_1}{\Delta'_2}; \Phi_3}{
              \Phi_1; \Sigma; \Gamma \vdash \s =_\beta \s_1 \cap \s_2
              &
              \Phi_1; \Sigma; \Gamma \vdash \Delta_1 : \s_1 \downrefine \Delta'_1; \Phi_2
              &
              \Phi_2; \Sigma; \Gamma \vdash \Delta_2 : \s_2 \downrefine \Delta'_2; \Phi_3
    }
    \\[1mm]
    \infer[(Proj_1)]{\Phi_1; \Sigma; \Gamma \vdash \prl \Delta : \s \downrefine \prl \Delta'; \Phi_3}{
      \Phi_1, (\Gamma \vdash ?x : \Type); \Sigma; \Gamma \vdash \s \cap ?x : \Type \downrefine \t; \Phi_2
      &
      \Phi_2; \Sigma; \Gamma \vdash \Delta : \s \cap ?x \downrefine \Delta'; \Phi_3
    }
    \\[1mm]
    \infer[(Proj_2)]{\Phi_1; \Sigma; \Gamma \vdash \prr \Delta : \s \downrefine \prr \Delta'; \Phi_3}{
      \Phi_1, (\Gamma \vdash ?x : \Type); \Sigma; \Gamma \vdash ?x \cap \s : \Type \downrefine \t; \Phi_2
      &
      \Phi_2; \Sigma; \Gamma \vdash \Delta : ?x \cap \s \downrefine \Delta'; \Phi_3
    }
    \\[1mm]
    \infer[(\mathit{Inj})]{\Phi_1; \Sigma; \Gamma \vdash \inni \s \Delta : \t \downrefine \inni \s' \Delta'; \Phi_3}{
      \Phi_1; \Sigma; \Gamma \vdash \t =_\beta \t_1 \cup \t_2
      & \Phi_1; \Sigma; \Gamma \vdash \s : \Type \downrefine \s'; \Phi_2
      & \Phi_2; \Sigma; \Gamma \vdash \s' \unif \t_i \unifresult \Phi_3
    }
    \\[1mm]
    \infer[(\mathit{Wildcard})]{\Phi; \Sigma; \Gamma \vdash \_ : \s \downrefine ?x[\metastrip{\Gamma}]; \Phi, (\Gamma \vdash ?x : \s)}{}
  \end{array}
  $
  \vspace{-3mm}
\caption{Rules for $\protect\downrefine$} \label{downrefine}
  \vspace{-3mm}
\end{figure}

\begin{figure}[t!]
\hspace{-2cm}
  $\begin{array}{c}
    \infer[(\mathit{Spair})]{\Phi_1; \Sigma; \Psi \vdash \spair{\Delta_1}{\Delta_2} \upessence M; \Phi_3}{
    \Phi_1; \Sigma; \Psi \vdash \Delta_1 \upessence M; \Phi_2
    & \Phi_2; \Sigma; \Psi \vdash M \loc \Delta_2 \downessence \Phi_3
      }
    \qquad 
    \infer[(\mathit{Proj})]{\Phi_1; \Sigma; \Psi \vdash \pri \Delta \upessence M; \Phi_2}{
    \Phi_1; \Sigma; \Psi \vdash \Delta \upessence M; \Phi_2
    }
    \\[1mm]
    \infer[(\mathit{Ssum})]{\Phi_1; \Sigma; \Psi \vdash \match{\Delta}{\s}{x\of\s_1}{\Delta_1}{x\of\s_2}{\Delta_2} \upessence (\lambda x.M)\at N; \Phi_7}{
      \begin{array}{l@{\qquad}l@{\qquad}l}
      \Phi_1; \Sigma; \Psi \vdash \Delta \upessence N; \Phi_2 &
      \Phi_2; \Sigma; \Psi \vdash \s \upessence \varsigma; \Phi_3 & 
      \Phi_3; \Sigma; \Psi \vdash \s_1 \upessence \varsigma_1; \Phi_4 \\[0mm] 
      \Phi_4; \Sigma; \Psi, x \vdash \Delta_1 \upessence M ; \Phi_5 & 
      \Phi_5; \Sigma; \Psi \vdash \s_2 \upessence \varsigma_2; \Phi_6 &
      \Phi_6; \Sigma; \Psi, x \vdash M \loc \Delta_2 \downessence \Phi_7
      \end{array}
    }
    \\[1mm]
    \infer[(\mathit{Inj})]{\Phi_1; \Sigma; \Psi \vdash \inni \s\ \Delta \upessence M; \Phi_2}{
    \Phi_1; \Sigma; \Psi \vdash \Delta \upessence M; \Phi_2
    }
   \qquad  
    \infer[(\mathit{Coe})]{\Phi_1; \Sigma; \Psi \vdash \coe\at \s\at \Delta \upessence M; \Phi_2}{
    \Phi_1; \Sigma; \Psi \vdash \Delta \upessence M; \Phi_2
    }
    \\[1mm]
    \infer[(\mathit{Prod})]{\Phi_1; \Sigma; \Psi \vdash \Pi x\of\s_1.\s_2 \upessence \Pi x\of\varsigma_1. \varsigma_2; \Phi_3}{
    \Phi_1; \Sigma; \Psi \vdash \s_1 \upessence \varsigma_1; \Phi_2
    & \Phi_2; \Sigma; \Psi, x \vdash \s_2 \upessence \varsigma_2; \Phi_3
      }
    \qquad 
    \infer[(\mathit{Abs})]{\Phi_1; \Sigma; \Psi \vdash \lambda x\of\s.\Delta \upessence \lambda x. M; \Phi_3}{
    \Phi_1; \Sigma; \Psi \vdash \s \upessence \varsigma; \Phi_2
    & \Phi_2; \Sigma; \Psi, x \vdash \Delta \upessence M; \Phi_3
      }
    \\[1mm]
    \infer[(App_1)]{\Phi_1; \Sigma; \Psi \vdash \Delta\at () \upessence M; \Phi_2}{
      \Phi_1; \Sigma; \Psi \vdash \Delta \upessence M; \Phi_2
    }
    \qquad 
    \infer[(App_2)]{\Phi_1; \Sigma; \Psi \vdash \Delta_1\at (S; \Delta_2) \upessence M\at N; \Phi_3}{
      \Phi_1; \Sigma; \Psi \vdash \Delta_1\at S \upessence M; \Phi_2 &
      \Phi_1; \Sigma; \Psi \vdash \Delta_2 \upessence N; \Phi_3
    }
    \\[1mm]
    \infer[(\cap)]{\Phi_1; \Sigma; \Psi \vdash \s_1 \cap \s_2 \upessence \varsigma_1 \cap \varsigma_2; \Phi_3}{
    \Phi_1; \Sigma; \Psi \vdash \s_1 \upessence \varsigma_1; \Phi_2
    & \Phi_2; \Sigma; \Psi \vdash \s_2 \upessence \varsigma_2; \Phi_3
      }
    \qquad 
    \infer[(\cup)]{\Phi_1; \Sigma; \Psi \vdash \s_1 \cup \s_2 \upessence \varsigma_1 \cup \varsigma_2; \Phi_3}{
    \Phi_1; \Sigma; \Psi \vdash \s_1 \upessence \varsigma_1; \Phi_2
    & \Phi_2; \Sigma; \Psi \vdash \s_2 \upessence \varsigma_2; \Phi_3
    }
  \end{array}
  $
  \vspace{-3mm}
\caption{Rules for $\protect\upessence$} \label{upessence}
  \vspace{-3mm}
\end{figure}

\begin{figure}[t!]
  $\begin{array}{c}
    \infer[(\mathit{Default})]{\Phi_1; \Sigma; \Psi \vdash M_1 \loc \Delta \downessence \Phi_3}{
    \Phi_1; \Sigma; \Psi \vdash \Delta \upessence M_2;\Phi_2
    & \Phi_2; \Sigma; \Psi \vdash M_1 \unif M_2 \unifresult \Phi_3
      }
    \\[1mm]
    \infer[(\mathit{Spair})]{\Phi_1; \Sigma; \Psi \vdash M \loc \spair{\Delta_1}{\Delta_2} \downessence \Phi_3}{
      \Phi_1; \Sigma; \Psi \vdash M \loc \Delta_1 \downessence \Phi_2
      & \Phi_2; \Sigma; \Psi \vdash M \loc \Delta_1 \downessence \Phi_3
    }
    \\[1mm]
    \infer[(\mathit{Proj})]{\Phi_1; \Sigma; \Psi \vdash M \loc \pri \Delta \downessence \Phi_2}{
      \Phi_1; \Sigma; \Psi \vdash M \loc \Delta\downessence \Phi_2
    }
    \qquad 
    \infer[(\mathit{Inj})]{\Phi_1; \Sigma; \Psi \vdash M \loc \inni \s\at\Delta \downessence \Phi_3}{
      \Phi_1; \Sigma; \Psi \vdash \sigma \upessence \varsigma; \Phi_2
      & \Phi_2; \Sigma; \Psi \vdash M \loc \Delta \downessence; \Phi_3
    }
    \\[1mm]
    \infer[(\mathit{Let})]{\Phi_1; \Sigma; \Psi \vdash M \loc \llet{x}{\s}{\Delta_1}{\Delta_2} \downessence \Phi_4}{
      \Phi_1; \Sigma; \Psi \vdash \s \upessence \varsigma; \Phi_2
      & \Phi_2; \Sigma; \Psi \vdash \Delta_1 \upessence M_1; \Phi_3
      & \Phi_3; \Sigma; \Psi, x := M_1 \vdash M \loc \Delta_2 \downessence \Phi_4
    }
    \\[1mm]
    \infer[(\mathit{Prod})]{\Phi_1; \Sigma; \Psi \vdash M \loc \Pi x\of\s_1.\s_2 \downessence \Phi_3}{
      \Phi_1; \Sigma; \Psi \vdash M =_\beta \Pi x\of\varsigma_1. \varsigma_2
      & \Phi_1; \Sigma; \Psi \vdash \varsigma_1 \loc \s_1 \downessence \Phi_2
      & \Phi_2; \Sigma; \Psi, x \vdash \varsigma_2 \loc \s_2 \downessence \Phi_3
    }
    \\[1mm]
    \infer[(\mathit{Abs})]{\Phi_1; \Sigma; \Psi \vdash M_1 \loc \lambda x\of\s.\Delta \downessence \Phi_2}{
      \Phi_1; \Sigma; \Psi \vdash M_1 =_\beta \lambda x. M_2
      & \Phi_1; \Sigma; \Psi, x \vdash M_2 \loc \Delta \downessence \Phi_2
    }
    \\[1mm]
    \infer[(\cap)]{\Phi_1; \Sigma; \Psi \vdash M \loc \s_1 \cap \s_2 \downessence \Phi_3}{
      \Phi_1; \Sigma; \Psi \vdash M =_\beta \varsigma_1 \cap \varsigma_2
      & \Phi_1; \Sigma; \Psi \vdash \varsigma_1 \loc \s_1 \downessence \Phi_2
      & \Phi_2; \Sigma; \Psi \vdash \varsigma_2 \loc \s_2 \downessence \Phi_3
    }
    \\[1mm]
    \infer[(\cup)]{\Phi_1; \Sigma; \Psi \vdash M \loc \s_1 \cup \s_2 \downessence \Phi_3}{
      \Phi_1; \Sigma; \Psi \vdash M =_\beta \varsigma_1 \cup \varsigma_2
      & \Phi_1; \Sigma; \Psi \vdash \varsigma_1 \loc \s_1 \downessence \Phi_2
      & \Phi_2; \Sigma; \Psi \vdash \varsigma_2 \loc \s_2 \downessence \Phi_3
    }
  \end{array}
  $
  \vspace{-3mm}
\caption{Rules for $\protect\downessence$} \label{downessence}
  \vspace{-3mm}
\end{figure}

\vspace{-3mm}
\section{The Read-Eval-Print-Loop of {\color{blue} \href{https://github.com/cstolze/Bull}{Bull}}}\label{sec:repl}
\vspace{-3mm}
The \emph{Read-Eval-Print-Loop} (REPL) reads a command which is given by the parser as a list of atomic commands. For instance, if the user writes:
\begin{bull}
Axiom (a b : Type) (f : a -> b).
\end{bull}
The parser creates the following list of three atomic commands:
\begin{enumerate}
\item the command asking \verb|a| to be an axiom of type \verb|Type|;
\item the command asking \verb|b| to be an axiom of type \verb|Type|;
\item the command asking \verb|f| to be an axiom of type \verb|a -> b|.
\end{enumerate}
The REPL tries to process the whole list. If there is a single failure while processing the list of atomic commands, it backtracks so the whole commands fails without changing the environment.
\noindent These commands are similar to the vernacular Coq commands and are quite intuitive. Here is the list of the REPL commands, along with their description:
\begin{bull}
Help.                               show this list of commands
Load "file".                        for loading a script file
Axiom term : type.                  define a constant or an axiom
Definition name [: type] := term.   define a term
Print name.                         print the definition of name
Printall.                           print all the signature
                                    (axioms and definitions)
Compute name.                       normalize name and print the result
Quit.                               quit
\end{bull}

\vspace{-3mm}
\section{Future work}\label{sec:afterbull}
\vspace{-3mm}
The current version of of {\color{blue} \href{https://github.com/cstolze/Bull}{Bull}} \cite{Bull} (ver. 1.0, December 2019) lacks of the following features that we plan to implement it in the next future.
\begin{enumerate}
\item \emph{Inductive types} are the most important feature to add, in order to have a usable theorem prover. We plan to take inspiration from the works of Paulin-Mohring \cite{paulinCIC}. This should be reasonably feasible;
\item \emph{Mixing subtyping and unification} is a difficult problem, especially with intersection and union types. The most extensive research which has been done in this domain is the work of Dudenhefner, Martens, and Rehof \cite{unif-inter}, where the authors study unification modulo subtyping with intersection types (but no union). It would be challenging to find a unification algorithm modulo subtyping for intersection and union types, but ideally it would allow us to do some implicit coercions. Take for example the famous Pierce code exploiting union and intersection types (full details in the distribution and in Appendix \ref{pierce-bull}): it would be interesting for the user to use implicit coercions in this way:
\begin{bull}
Axiom (Neg Zero Pos T F : Type) (Test : Pos | Neg).
Axiom Is_0 : (Neg -> F) & (Zero -> T) & (Pos -> F).
Definition Is_0_Test : F := smatch Test with
                                          x => coe _ Is_0 x
                                        , x => coe _ Is_0 x
                                        end.
\end{bull}
The unification algorithm would then guess that the first wildcard should be replaced with \verb|Pos -> F| and the second one should be replaced with \verb|Neg -> F|, which does not seem feasible if the unification algorithm does not take subtyping into account;

\item \emph{Relevant arrow}, as defined in \cite{FSTTCS18}, it could be useful to add more expressivity to our system. Relevant implication allows for a natural introduction of subtyping, in that $A \supset_{\sf r} B$ morally means $A \leqslant B $. Relevant implication amounts to a notion of ``proof-reuse''. Combining the remarks in \cite{BM94,BDdL}, minimal relevant implication, strong intersection and strong union correspond respectively to the implication, conjunction and disjunction operators of Meyer and Routley's Minimal Relevant Logic $B^+$ \cite{RM72}. This could lead to some implementation problem, because deciding $\beta$-equality for the essences in this extended system would be undecidable;
\item A \emph{Tactic language}, such as the one of Coq, should be useful. Currently, there is no such tactic language for {\color{blue} \href{https://github.com/cstolze/Bull}{Bull}}: conceiving such a language should be feasible in the medium term.
\end{enumerate}

\noindent {\bf Acknowlegements.} This work could not be have be done without the many useful discussions with Furio Honsell, Ivan Scagnetto, Ugo de'Liguoro, and Daniel Dougherty.

\bibliography{inter_biblio}

\newpage
\appendix
\noindent {\LARGE \bf Appendix at the discretion of the reviewers}
\vspace{-3mm}
\section{Examples}
\vspace{-3mm}



This appendix show examples which show a uniform and approach to the encoding of a plethora of type disciplines and systems which ultimately stem or can capitalize from strong proof-functional connectives and subtyping. The framework $\DLF$ presented in \cite{FSTTCS18}, and its software incarnation {\color{blue} \href{https://github.com/cstolze/Bull}{Bull}} introduced in this paper is the first to accommodate all the examples and counterexamples that have appeared in the literature.

The Appendix is organized as follows: in Subsection \ref{encodings} we present some examples in $\DLF$ along with their code in {\color{blue} \href{https://github.com/cstolze/Bull}{Bull}}, and in Subsection \ref{coq-encodings}, we show some similar encodings done done in LF\footnote{For convenience, we wrote and typechecked these examples in Coq, as LF is a sublanguage of the Calculus of Constructions.}, in order to emphasize the benefits of $\DLF$.  In what follows, we denote by $\B$ \cite{BDdL} the union and intersection type assignment system \ala\ Curry\footnote{Type inference is, of course, undecidable.}.

\vspace{-3mm}
\subsection{\texorpdfstring{Encodings in $\DLF$}{Encodings in LF-Delta}}\label{encodings}
\vspace{-3mm}
We start by showing the expressive power of $\DLF$ in encoding classical features of typing disciplines with strong intersection and union. For these examples, we set $\Sigma \eqdef \s\of\Type, \t\of\Type$.

\medskip
\noindent {\bf Auto application.} The judgment $\vdash_{\B}\lambda x. x \at x :\s \cap (\s \to \t) \to \t$ in $\B$, is rendered in $\DLF$ by the $\DLF$ judgment $ \sigmadash \lambda x \of\s \cap (\s \to \t). (\prr x) \at (\prl x): \s \cap (\s \to \t) \to \t$.

\medskip
\noindent {\bf Polymorphic identity.} The judgment $\vdash_{\B}\lambda x.x:(\s \to \s) \cap (\t \to \t)$ in $\B$, is rendered in $\DLF$ by the judgment
$\vdash_{\Sigma}
\spair{\lambda x \of \s.x}{\lambda x\of \t.x}: (\s \to \s) \cap (\t \to \t)
$.

\medskip
\noindent {\bf Commutativity of union.} The judgment $\vdash_{\Sigma} \lambda x.x :(\s \cup \t) \to (\t \cup \s)$ in $\B$, is rendered in $\DLF$ by the judgment
$\vdash_{\Sigma}
\lambda x\of\s{\cup}\t. \ssum{\lambda y\of\s. \inr{\t} y}{\lambda y\of\t. \inl{\s} y} \at x :(\s \cup \t) \to (\t \cup \s).
$

\medskip
\noindent
The {\color{blue} \href{https://github.com/cstolze/Bull}{Bull}} code corresponding to these examples is the following:
\begin{bull}
Axiom (s t : Type).
Definition auto_application (x : s & (s -> t)) := (proj_r x) (proj_l x).
Definition poly_id : (s -> s) & (t -> t) := let id1 x := x in
                                         let id2 x := x in < id1, id2 >.
Definition commut_union (x : s | t) := smatch x with
                                                  x : s => inj_r t x
                                                , x : t => inj_l s x
                                                end.
\end{bull}

\vspace{-3mm}
{Pierce's code}
\vspace{-3mm}
We recall the Pierce code showing expressivity of union and intersection types:
\begin{displaymath}
\begin{array}{rcl}
\mbox{Test}
&\eqdef &
\mbox{if } b
\mbox{ then } 1 \mbox{ else} -\!1 : Pos \cup Neg
\\[1mm]
\mbox{Is}\_0
& : &
(Neg \to F) \cap
(Zero \to T) \cap
(Pos \to F)
\\[1mm]
(\mbox{Is}\_0 \mbox{ Test}) &:& F
\end{array}
\end{displaymath}
\noindent The expressive power of union types highlighted by Pierce is rendered in $\DLF$ by:
\begin{displaymath}
\begin{array}{rcl}
    Neg &:&  \Type \quad Zero : \Type \quad Pos : \Type \quad
    T : \Type \quad F : \Type \quad
    \mbox{Test} : Pos \cup Neg
    \\[1mm]
   \mbox{Is}\_0 & : & (Neg \rightarrow F) \cap ((Zero \rightarrow T) \cap (Pos \rightarrow F)) \\[1mm]
  \mbox{Is}\_0\_\mbox{Test} & \eqdef &
  \ssum{\lambda x \of Pos. (\prr \prr \mbox{Is}\_0) \at x}{\lambda x \of Neg. (\prl \mbox{Is}\_0) \at x} \at \mbox{Test}
\end{array}
\end{displaymath}
\noindent The {\color{blue} \href{https://github.com/cstolze/Bull}{Bull}} code corresponding to this example is the following:\label{pierce-bull}
\begin{bull}
Axiom (Neg Zero Pos T F : Type) (Test : Pos | Neg).
Axiom Is_0 : (Neg -> F) & (Zero -> T) & (Pos -> F).
Definition Is_0_Test := smatch Test with
                                      x => coe (Pos -> F) Is_0 x
                                    , x => coe (Neg -> F) Is_0 x
                                    end.
\end{bull}
As you can see, the code is quite short and readable, in contrast to the LF encoding of the same example found in Figure \ref{pierce-coq}.

\vspace{-3mm}
\subsection{Hereditary Harrop formul{\ae}}
\vspace{-3mm}
\begin{figure}[t]{
    \small
    \begin{displaymath}
      \begin{array}{lcl}
        \multicolumn{3}{l}{
          \mbox{Atomic propositions, non-atomic goals and non-atomic programs:}
          \quad
          \alpha,\gamma_{0},\pi_{0} : \Type} \\

        \multicolumn{3}{l}{ \mbox{Goals and programs:}\quad
          \gamma = \alpha \cup \gamma_{0} \qquad \pi = \alpha \cup \pi_{0}} \\

        \multicolumn{3}{l}{
          \mbox{Constructors (implication, conjunction, disjunction).}} \\[3mm]
        \mathsf{impl} & : & (\pi \rightarrow \gamma \rightarrow \gamma_{0}) \cap (\gamma \rightarrow \pi \rightarrow \pi_{0}) \\
        \mathsf{impl}_1 & = & \lambda x \of \pi. \lambda y \of \gamma. \inr{\alpha}(\prl \mathsf{impl}\at x\at y)
        \qquad \mathsf{impl}_2 = \lambda x \of \gamma. \lambda y \of \pi. \inr{\alpha}(\prr \mathsf{impl}\at x\at y) \\
        \mathsf{and} & : & (\gamma \rightarrow \gamma \rightarrow \gamma_{0}) \cap (\pi \rightarrow \pi \rightarrow \pi_{0})\\
        \mathsf{and}_1 & =& \lambda x \of \gamma. \lambda y \of \gamma. \inr{\alpha} (\prl \mathsf{and}\at x\at y)
        \qquad\,\, \mathsf{and}_2 = \lambda x \of \pi. \lambda y \of \pi. \inr{\alpha} (\prr \mathsf{and}\at x\at y)\\
        \mathsf{or} & : & (\gamma \rightarrow \gamma \rightarrow \gamma_{0})
        \qquad \mathsf{or}_1 = \lambda x \of \gamma. \lambda y \of \gamma. \inr{\alpha} (\mathsf{or}\at x\at y) \\

        \multicolumn{3}{l}{
          \mbox{$\mathsf{solve}\at p\at g$ indicates that the judgment $p \vdash g$ is valid.}}\\

        \multicolumn{3}{l}{
          \mbox{$\mathsf{bchain}\at p\at a\at g$ indicates that, if $p \vdash g$ is valid, then $p \vdash a$ is valid.}}\\
        \mathsf{solve} & : & \pi \rightarrow \gamma \rightarrow \Type
        \qquad
        \mathsf{bchain} : \pi \rightarrow \alpha \rightarrow \gamma \rightarrow \Type \\

        \multicolumn{3}{l}{
          \mbox{Rules for $\mathsf{solve}$:}} \\
        - & : & \Pi (p \of \pi) (g_{1}, g_{2} \of \gamma). \mathsf{solve}\at p\at g_{1} \rightarrow \mathsf{solve}\at p\at g_{2} \rightarrow \mathsf{solve}\at p\at (\mathsf{and}_1\at g_{1}\at g_{2}) \\
        - & : & \Pi (p \of \pi) (g_{1}, g_{2} \of \gamma). \mathsf{solve}\at p\at g_{1} \rightarrow \mathsf{solve}\at p\at (\mathsf{or}_1\at g_{1}\at g_{2}) \\
        - & : & \Pi (p \of \pi) (g_{1}, g_{2} \of \gamma). \mathsf{solve}\at p\at g_{2} \rightarrow \mathsf{solve}\at p\at (\mathsf{or}_1\at g_{1}\at g_{2}) \\
        - & : & \Pi (p_{1}, p_{2} \of \pi) (g \of \gamma). \mathsf{solve}\at (\mathsf{and}_2\at p_{1}\at p_{2})\at g \rightarrow \mathsf{solve}\at p_{1}\at (\mathsf{impl}_1\at p_{2}\at g) \\
        - & : & \Pi (p \of \pi)(a \of \alpha)(g \of \gamma). \mathsf{bchain}\at p\at a\at g \rightarrow \mathsf{solve}\at p\at g \rightarrow \mathsf{solve}\at p\at (\inl{\gamma_0} a) \\

        \multicolumn{3}{l}{
          \mbox{Rules for $\mathsf{bchain}$:}} \\
        - & : & \Pi (a \of \alpha)(g \of \gamma). \mathsf{bchain}\at (\mathsf{impl}_2\at g\at (\inl{\pi_0} a))\at a\at g \\
        - & : & \Pi (p_{1}, p_{2} \of \pi)(a \of \alpha)(g \of \gamma). \mathsf{bchain}\at p_{1}\at a\at g \rightarrow \mathsf{bchain}\at (\mathsf{and}_2\at p_{1}\at p_{2})\at a\at g \\
        - & : & \Pi (p_{1}, p_{2} \of \pi)(a \of \alpha)(g \of \gamma). \mathsf{bchain}\at p_{2}\at a\at g \rightarrow \mathsf{bchain}\at (\mathsf{and}_2\at p_{1}\at p_{2})\at a\at g \\
        - & : & \Pi (p \of \pi)(a \of \alpha)(g, g_{1}, g_{2} \of \gamma). \mathsf{bchain}\\
        & & (\mathsf{impl}_2\at (\mathsf{and}_1\at g_{1}\at g_{2})\at p)\at a\at g \rightarrow \mathsf{bchain}\at (\mathsf{impl}_2\at g_{1}\at (\mathsf{impl}_2\at g_{2}\at p))\at a\at g \\
        - & : & \Pi (p_{1}, p_{2} \of \pi)(a \of \alpha)(g, g_{1} \of \gamma). \mathsf{bchain}\at (\mathsf{impl}_2\at g_{1}\at p_{1})\at a\at g \rightarrow \mathsf{bchain}\\
        & & (\mathsf{impl}_2\at g_{1}\at (\mathsf{and}_2\at p_{1}\at p_{2}))\at a\at g \\
        - & : & \Pi (p_{1}, p_{2} \of \pi)(a \of \alpha)(g, g_{1} \of \gamma). \mathsf{bchain}\\
        & & (\mathsf{impl}_2\at g_{1}\at p_{2})\at a\at g \rightarrow \mathsf{bchain}\at (\mathsf{impl}_2\at g_{1}\at (\mathsf{and}_2\at p_{1}\at p_{2}))\at a\at g
      \end{array}
    \end{displaymath}
  }
  \vspace{-3mm}
  \caption{The $\DLF$ encoding of Hereditary Harrop Formul{\ae}}
   \vspace{-3mm}
 \label{HHF}
\end{figure}

The encoding of Hereditary Harrop's Formul{\ae} is one of the motivating examples given by Pfenning for introducing Refinement Types in LF \cite{Refine93,PfenningLovas-10}. In $\DLF$ it can be expressed as in Figure \ref{HHF}  and type checked in {\color{blue} \href{https://github.com/cstolze/Bull}{Bull}}, without any reference to intersection types, by a subtle use of union types. We add also rules for solving and backchaining. Hereditary Harrop formul{\ae} can be recursively defined using two mutually recursive syntactical objects called programs ($\pi$) and goals ($\gamma$):
$$\begin{array}{rcl@{\qquad\qquad}rcl}
  \gamma & := & \alpha \mid \gamma \wedge \gamma \mid \pi \Rightarrow \gamma \mid \gamma \vee \gamma
&
\pi & := & \alpha \mid \pi \wedge \pi \mid \gamma \Rightarrow \pi
\end{array}
$$
We can provide an alternative encoding of atoms, goals and programs which is more faithful to the one by Pfenning. Namely, we can introduce in the signature the constants $c_1:\alpha\rightarrow^{\sf r}\gamma$ and $c_2:\alpha\rightarrow^{\sf r}\pi$ in order to represent the axioms $atom\leq goal$ and $atom\leq prog$ in Pfenning's encoding. Our approach based on union types, while retaining the same expressivity permits to shortcut certain inclusions and to rule out also certain exotic goals and exotic programs. Indeed, for the purpose of establishing the adequacy of the encoding, it is sufficient to avoid variables involving union types in the derivation contexts.

\noindent The {\color{blue} \href{https://github.com/cstolze/Bull}{Bull}} code is the following:
\begin{bull}
(* three base types: atomic propositions, non-atomic goals and non-atomic programs *)
Axiom atom : Type.
Axiom non_atomic_goal : Type.
Axiom non_atomic_prog : Type.

(* goals and programs are defined from the base types *)
Definition goal := atom | non_atomic_goal.
Definition prog := atom | non_atomic_prog.

(* constructors (implication, conjunction, disjunction) *)
Axiom impl : (prog -> goal -> non_atomic_goal) & (goal -> prog -> non_atomic_prog).
Definition impl_1 p g := inj_r atom (proj_l impl p g).
Definition impl_2 g p := inj_r atom (proj_r impl g p).
Axiom and : (goal -> goal -> non_atomic_goal) & (prog -> prog -> non_atomic_prog).
Definition and_1 g1 g2 := inj_r atom (proj_l and g1 g2).
Definition and_2 p1 p2 := inj_r atom (proj_r and p1 p2).
Axiom or : (goal -> goal -> non_atomic_goal).
Definition or_1 g1 g2 := inj_r atom (or g1 g2).

(* solve p g means: the judgment p |- g is valid *)
Axiom solve : prog -> goal -> Type.

(* backchain p a g means: if p |- g is valid, then p |- a is valid *)
Axiom backchain : prog -> atom -> goal -> Type.

(* rules for solve *)
Axiom solve_and : forall p g1 g2, solve p g1 -> solve p g2 -> solve p (and_1 g1 g2).
Axiom solve_or1 : forall p g1 g2, solve p g1 -> solve p (or_1 g1 g2).
Axiom solve_or2 : forall p g1 g2, solve p g2 -> solve p (or_1 g1 g2).
Axiom solve_impl : forall p1 p2 g, solve (and_2 p1 p2) g -> solve p1 (impl_1 p2 g).
Axiom solve_atom : forall p a g, backchain p a g -> solve p g -> solve p (inj_l non_atomic_goal a).

(* rules for backchain *)
Axiom backchain_and1 :
 forall p1 p2 a g, backchain p1 a g -> backchain (and_2 p1 p2) a g.
Axiom backchain_and2 :
 forall p1 p2 a g, backchain p1 a g -> backchain (and_2 p1 p2) a g.
Axiom backchain_impl_atom :
 forall a g, backchain (impl_2 g (inj_l non_atomic_prog a)) a g.
Axiom backchain_impl_impl :
 forall p a g g1 g2, backchain (impl_2 (and_1 g1 g2) p) a g -> backchain (impl_2 g1 (impl_2 g2 p)) a g.
Axiom backchain_impl_and1 :
 forall p1 p2 a g g1, backchain (impl_2 g1 p1) a g -> backchain (impl_2 g1 (and_2 p1 p2)) a g.
Axiom backchain_impl_and2 :
 forall p1 p2 a g g1, backchain (impl_2 g1 p2) a g -> backchain (impl_2 g1 (and_2 p1 p2)) a g.
\end{bull}
\vspace{-3mm}
\subsection{Natural deductions in normal form}
\vspace{-3mm}
The second motivating example for intersection types given in~\cite{Refine93,PfenningLovas-10} is {\em natural deductions in normal form}. We recall that a natural deduction is in normal form if there are no applications of elimination rules of a logical connective immediately following their corresponding introduction, in the main branch of a subderivation.

\begin{displaymath}
  \begin{array}{rcl}
    o &:& Type\qquad
    \supset : o\to o\to o\qquad
    Elim, {\rm Nf}^0 : o\to\Type\qquad\\[3mm]
    {\rm Nf} & \equiv & \lambda A\of o.{\rm Nf}^0(A)\cup Elim(A)\\[3mm]
    \supset_I & : & \Pi A,B\of o.(Elim(A)\to {\rm Nf}(B))\to {\rm Nf}^0(A\supset B)\\[3mm]
    \supset_E & : & \Pi A,B\of o.Elim(A\supset B)\to {\rm Nf}^0(A)\to Elim(B).
  \end{array}
\end{displaymath}
The corresponding {\color{blue} \href{https://github.com/cstolze/Bull}{Bull}} code is the following:
\begin{bull}
Axiom (o : Type) (impl : o -> o -> o) (Elim Nf0 : o -> Type).
Definition Nf A := Nf0 A | Elim A.
Axiom impl_I : forall A B, (Elim A -> Nf B) -> (Nf0 (impl A B)).
Axiom impl_E : forall A B, Elim (impl A B) -> Nf0 A -> Elim B.
\end{bull}
The encoding we give in $\DLF$ is a slightly improved version of the one in~\cite{Refine93,PfenningLovas-10}: as Pfenning, we restrict to the purely implicational fragment. As in the previous example, we use union types to define normal forms ${\rm Nf}(A)$ either as pure elimination-deductions from hypotheses $Elim(A)$ or normal form-deductions ${\rm Nf}^0(A)$. As above we could have used also intersection types. This example is interesting in itself, being the prototype of the encoding of type systems using canonical and atomic syntactic categories \cite{harper-licata} and also of Fitch Set Theory \cite{Fitch-APLAS16}.

\smallskip \noindent \textbf{Metacircular Encodings.} This example uses an experimental implementation of relevant arrows in Bull.
The following diagram summarizes the network of adequate encodings/inclusions between $\DLF$, LF, and $\B$ that can be defined:
\begin{displaymath}
\xymatrix{
 \mathrm{LF}\ar@{=>}[r]^{sh} & \mathrm{LF}_\Delta\ar@{=>}[r]^{dp} & \mathrm{LF}\\ 
 \B \ar@{=>}[ru]^{sh}\ar@{=>}[r]^{dp} & \mathrm{LF}\ar@{^{(}->}[u] &
}
\end{displaymath}
We denote by ${\cal S}_1 \Longrightarrow {\cal S}_2$ the encoding of system ${\cal S}_1$ in system ${\cal S}_2$, where the label \textit{sh} (resp. \textit{dp}), denotes a shallow (resp. deep) embedding. The notation ${\cal S}_1\hookrightarrow {\cal S}_2$ denotes that ${\cal S}_2$ is an extension of ${\cal S}_1$.

With the intention of providing a better formal understanding of the semantics of strong intersection and union types in a logical framework, we provide in Section \ref{coq-encodings} a deep LF encoding of a presentation of $\B$ \ala\ Church \cite{APLAS16}.
An encoding of $\B$ in $\DLF$ can be mechanically type checked in the environment \cite{LFMTP17,FSTTCS18}. We even can add the relevant arrow. The corresponding {\color{blue} \href{https://github.com/cstolze/Bull}{Bull}} code is the following:
\begin{bull}
(* Object type *)
Axiom o : Type.

(* Type connectives *)
Axiom carrow : o -> o -> o.
Axiom cinter : o -> o -> o.
Axiom cunion : o -> o -> o.
Axiom crelev : o -> o -> o.

(* Transform the object types into real types *)
Axiom aOk : o -> Type.

(* Semantics *)
Axiom cabst    : forall s t, (aOk s -> aOk t) >> aOk (carrow s t).
Axiom capp     : forall s t, aOk (carrow s t) >> aOk s -> aOk t.
Axiom csabst   : forall s t,  (aOk s >> aOk t) >> aOk (crelev s t).
Axiom csapp    : forall s t, aOk (crelev s t) >> aOk s >> aOk t.
Axiom cpair    : forall s t, (aOk s & aOk t) >> aOk (cinter s t).
Axiom cpri     : forall s t, aOk (cinter s t) >> (aOk s & aOk t).
Axiom cini     : forall s t, (aOk s | aOk t) >> aOk (cunion s t).
Axiom ccopair : forall s t, aOk (cunion s t) >> (aOk s | aOk t).
\end{bull}
Using this encoding, we can deeply encode self-application $$\lambda x.x\at x : (\s \cap (\s \rightarrow \t)) \rightarrow \t$$ and commutativity of union $$\lambda x. x : (\s \cup \t) \rightarrow (\t \cup \s)$$
\begin{bull}
Axiom s : o.
Axiom t : o.
Definition halfomega :=
 cabst (cinter s (carrow s t)) t (fun x => capp s t (proj_r (cpri s (carrow s t) x))
  (proj_l (cpri s (carrow s t) x))).
Definition idpair :=
 cpair (carrow s s) (carrow t t) <cabst s s (fun x => x), cabst t t (fun x => x)>.
Definition communion := cabst (cunion s t) (cunion t s)
 (fun x => smatch ccopair s t x with
             y => cini t s (inj_r (aOk t) y)
           , y => cini t s (inj_l (aOk s) y)
           end).
\end{bull}
We also can show that the commutativity of union with a relevant arrow $\lambda x. x : (\s \cup \t) \rightarrow^{\sf r} (\t \cup \s)$:
\begin{bull}
Definition communion' := csabst (cunion s t) (cunion t s)
 (sfun x => smatch ccopair s t x with
             y => cini t s (inj_r (aOk t) y)
           , y => cini t s (inj_l (aOk s) y)
           end).
\end{bull}
A shallow encoding of LF in $\DLF$ making essential use of intersection types can be also type checked. The corresponding {\color{blue} \href{https://github.com/cstolze/Bull}{Bull}} code is the following:
\begin{bull}
Axiom obj' : Type.
Axiom fam' : Type.
Axiom knd' : Type.
Axiom sup' : Type.

Axiom same : obj' & fam' & knd' & sup'.
Axiom term : (obj' | fam' | knd' | sup') -> Type.
(* The obj, fam, knd, and sup types have the same essence (term same) *)
Definition obj := term (coe (obj' | fam' | knd' | sup') (coe obj' same)).
Definition fam := term (coe (obj' | fam' | knd' | sup') (coe fam' same)).
Definition knd := term (coe (obj' | fam' | knd' | sup') (coe knd' same)).
Definition sup := term (coe (obj' | fam' | knd' | sup') (coe sup' same)).

Axiom tp : knd & sup.
(* star and sqre have the same essence (tp) *)
Definition star := coe knd tp.
Definition sqre := coe sup tp.

Axiom lam : (fam -> (obj -> obj) -> obj) & (fam -> (obj -> fam) -> fam).
Definition lam_1 := coe (fam -> (obj -> obj) -> obj) lam.
Definition lam_2 := coe (fam -> (obj -> fam) -> fam) lam.

Axiom pi : (fam -> (obj -> fam) -> fam) & (fam -> (obj -> knd) -> knd).
Definition pi_1 := coe (fam -> (obj -> fam) -> fam) pi.
Definition pi_2 := coe (fam -> (obj -> knd) -> knd) pi.

Axiom app : (obj -> obj -> obj) & (fam -> obj -> fam).
Definition app_1 := coe (obj -> obj -> obj) app.
Definition app_2 := coe (fam -> obj -> fam) app.

Axiom of_1 : obj -> fam -> Type.
Axiom of_2 : fam -> knd -> Type.
Axiom of_3 : knd -> sup -> Type.

Axiom of_ax : of_3 star sqre.
(* Rules for lambda-abstraction are "essentially" the same *)
Definition of_lam1 := forall t1 t2 t3, of_2 t1 star ->
      (forall x, of_1 x t1 -> of_1 (t2 x) (t3 x)) -> of_1 (lam_1 t1 t2) (pi_1 t1 t3).
Definition of_lam2 := forall t1 t2 t3, of_2 t1 star ->
      (forall x, of_1 x t1 -> of_2 (t2 x) (t3 x)) -> of_2 (lam_2 t1 t2) (pi_2 t1 t3).
Axiom of_lam : of_lam1 & of_lam2.
(* Rules for product are ''essentially'' the same *)
Definition of_pi1 := forall t1 t2, of_2 t1 star ->
 (forall x, of_1 x t1 -> of_2 (t2 x) star) -> of_2 (pi_1 t1 t2) star.
Definition of_pi2 := forall t1 t2, of_2 t1 star ->
 (forall x, of_1 x t1 -> of_3 (t2 x) sqre) -> of_3 (pi_2 t1 t2) sqre.
Axiom of_pi : of_pi1 & of_pi2.
(* Rules for application are ''essentially'' the same *)
Definition of_app1 := forall t1 t2 t3 t4, of_1 t1 (pi_1 t3 t4) ->
 of_1 t2 t3 -> of_1 (app_1 t1 t2) (t4 t2).
Definition of_app2 := forall t1 t2 t3 t4, of_2 t1 (pi_2 t3 t4) ->
 of_1 t2 t3 -> of_2 (app_2 t1 t2) (t4 t2).
Axiom of_app : of_app1 & of_app2.
\end{bull}
\noindent We finish this chapter by providing examples of encoding in LF.
\vspace{-3mm}
\section{Encodings in LF}\label{coq-encodings}
\vspace{-3mm}
We present a pure LF encoding of a presentation of $\B$ \ala\ Church, using the Coq syntax, and the \emph{Higher-Order Abstract Syntax} (HOAS) \cite{hoas}. We use HOAS in order to take advantage of the higher-order features of the frameworks: other abstract syntax representation techniques would not be much different, but more verbose:
\begin{coq}
(* Define our types *)
Axiom o : Set.
(* Axiom omegatype : o. *)
Axioms (arrow inter union : o -> o -> o).

(* Transform our types into LF types *)
Axiom OK : o -> Set.

(* Define the essence equality as an equivalence relation *)
Axiom Eq : forall (s t : o), OK s -> OK t -> Prop.
Axiom Eqrefl : forall (s : o) (M : OK s), Eq s s M M.
Axiom Eqsymm : forall (s t : o) (M : OK s) (N : OK t), Eq s t M N -> Eq t s N M.
Axiom Eqtrans : forall (s t u : o) (M : OK s) (N : OK t) (O : OK u),
 Eq s t M N -> Eq t u N O -> Eq s u M O.

(* constructors for arrow (->I and ->E) *)
Axiom Abst : forall (s t : o), ((OK s) -> (OK t)) -> OK (arrow s t).
Axiom App : forall (s t : o), OK (arrow s t) -> OK s -> OK t.

(* constructors for intersection *)
Axiom Proj_l : forall (s t : o), OK (inter s t) -> OK s.
Axiom Proj_r : forall (s t : o), OK (inter s t) -> OK t.
Axiom Pair : forall (s t : o) (M : OK s) (N : OK t), Eq s t M N -> OK (inter s t).

(* constructors for union *)
Axiom Inj_l : forall (s t : o), OK s -> OK (union s t).
Axiom Inj_r : forall (s t : o), OK t -> OK (union s t).
Axiom Sum : forall (s t u : o) (X : OK (arrow s u)) (Y : OK (arrow t u)),
 OK (union s t) -> Eq (arrow s u) (arrow t u) X Y -> OK u.

(* define equality wrt arrow constructors *)
Axiom Eqabst : forall (s t s' t' : o) (M : OK s -> OK t) (N : OK s' -> OK t'),
(forall (x : OK s) (y : OK s'), Eq s s' x y -> Eq t t' (M x) (N y)) ->
Eq (arrow s t) (arrow s' t') (Abst s t M) (Abst s' t' N).
Axiom Eqapp : forall (s t s' t' : o) (M : OK (arrow s t)) (N : OK s)
 (M' : OK (arrow s' t')) (N' : OK s'), Eq (arrow s t) (arrow s' t') M M' ->
 Eq s s' N N' -> Eq t t' (App s t M N) (App s' t' M' N').

(* define equality wrt intersection constructors *)
Axiom Eqpair : forall (s t : o) (M : OK s) (N : OK t) (pf : Eq s t M N),
 Eq (inter s t) s (Pair s t M N pf) M.
Axiom Eqproj_l : forall (s t : o) (M : OK (inter s t)),
 Eq (inter s t) s M (Proj_l s t M).
Axiom Eqproj_r : forall (s t : o) (M : OK (inter s t)),
 Eq (inter s t) t M (Proj_r s t M).

(* define equality wrt union *)
Axiom Eqinj_l : forall (s t : o) (M : OK s), Eq (union s t) s (Inj_l s t M) M.
Axiom Eqinj_r : forall (s t : o) (M : OK t), Eq (union s t) t (Inj_r s t M) M.
Axiom Eqsum : forall (s t u : o) (M : OK (arrow s u)) (N : OK (arrow t u))
 (O : OK (union s t)) (pf: Eq (arrow s u) (arrow t u) M N) (x : OK s),
 Eq s (union s t) x O -> Eq u u (App s u M x) (Sum s t u M N O pf).
\end{coq}
The \texttt{Eq} predicate plays the same role of the essence function in $\DLF$, namely, it encodes the judgment that two proofs (\ie two terms of type \texttt{(OK \_)}) have the same structure. This is crucial in the \texttt{Pair} axiom (\ie the introduction rule of the intersection type constructor) where we can inhabit the type \texttt{(inter s t)} only when the proofs of its component types \texttt{s} and \texttt{t} share the same structure (\ie we have a witness of type \texttt{(Eq s t M N)}, where \texttt{M} has type \texttt{(OK s)} and \texttt{N} has type \texttt{(OK t)}). A similar role is played by the \texttt{Eq} premise in the \texttt{Sum} axiom (\ie the elimination rule of the union type constructor). We have an \texttt{Eq} axiom for each proof rule.

Using this encoding, we can encode auto-application, polymorphic identity, and commutativity of union:
\begin{coq}
Section Examples.
  Hypotheses s t : o.

  (* lambda x. x x : (sigma inter (sigma -> tau)) -> tau *)
  Definition autoapp : OK (arrow (inter s (arrow s t)) t) :=
    Abst (inter s (arrow s t)) t (fun x : OK (inter s (arrow s t)) =>
     App s t (Proj_r s (arrow s t) x) (Proj_l s (arrow s t) x)).

  (* lambda x. x : (sigma -> sigma) inter (tau -> tau) *)
  Definition polyid : OK (inter (arrow s s) (arrow t t)) :=
    Pair (arrow s s) (arrow t t) (Abst s s (fun x : OK s => x))
     (Abst t t (fun x : OK t => x))
     (Eqabst s s t t (fun x : OK s => x) (fun x : OK t => x)
      (fun (x : OK s) (y : OK t) (Z : Eq s t x y) => Z)).

  (* lambda x. x : (sigma union tau) -> (tau union sigma) *)
  Definition commutunion : OK (arrow (union s t) (union t s)) :=
    Abst (union s t) (union t s)
         (fun x : OK (union s t) =>
            Copair s t (union t s) (Abst s (union t s) (fun y : OK s => Inj_r t s y))
                   (Abst t (union t s) (fun y : OK t => Inj_l t s y)) x
                   (Eqabst s (union t s) t (union t s) (fun y : OK s => Inj_r t s y)
                           (fun y : OK t => Inj_l t s y)
                           (fun (x0 : OK s) (y : OK t) (pf : Eq s t x0 y) =>
                              Eqtrans (union t s) s (union t s) (Inj_r t s x0) x0
                                      (Inj_l t s y)
                                      (Eqinj_r t s x0)
                                      (Eqtrans s t (union t s) x0 y (Inj_l t s y) pf
                                               (Eqsymm (union t s) t (Inj_l t s y) y
                                                       (Eqinj_l t s y)))))).
\end{coq}
The definition of \coqe|commutunion| is quite unreadable, and has been created from the following Ltac script:
\begin{coq}
  Definition commutunion' : OK (arrow (union s t) (union t s)).
  Proof.
    apply (Abst (union s t) (union t s)).
    intro x.
    apply (Copair _ _ _ (Abst _ _ (fun y : _ => Inj_r _ _ y))
                  (Abst _ _ (fun y : _ => Inj_l _ _ y)) x).
    apply Eqabst.
    intros x0 y pf.
    assert (Eq _ _ (Inj_r t _ x0) x0) by apply Eqinj_r.
    assert (Eq _ _ y (Inj_l _ s y)). {
      apply Eqsymm.
      apply Eqinj_l.
    }
    eapply Eqtrans.
    now apply H.
    eapply Eqtrans.
    now apply pf.
    trivial.
  Defined.
End Examples.
\end{coq}
Using the same encoding of $\DLF$ in Coq, the Pierce's code would be encoded as:
\begin{coq}
Section Test.
  Hypotheses (Pos Zero Neg T F : o).
  Hypotheses (Test : OK (union Pos Neg))
             (is_0 : OK (inter (arrow Neg F) (inter (arrow Zero T) (arrow Pos F)))).

  (* is_0 Test *)
  Definition is0test : OK F.
    apply (Copair _ _ _ (Abst _ _ (fun x : _ => App _ _ (Proj_r _ _ (Proj_r _ _ is_0)) x))
                  (Abst _ _ (fun x : _ => App _ _ (Proj_l _ _ is_0) x))).
    now apply Test.
    apply Eqabst.
    intros x y pf.
    apply Eqapp.
    - assert (H : Eq _ _ is_0 (Proj_r (arrow Neg F) (inter (arrow Zero T)
                                                            $\,\,$(arrow Pos F)) is_0))
        by apply Eqproj_r.
      assert (H0 : Eq _ _ (Proj_r (arrow Neg F) (inter (arrow Zero T)
                                                        (arrow Pos F)) is_0)
                      (Proj_r (arrow Zero T) (arrow Pos F)
                              (Proj_r (arrow Neg F) (inter (arrow Zero T)
                                                            $\,\,$(arrow Pos F)) is_0)))
        by apply Eqproj_r.
      assert (H1 : Eq _ _ is_0 (Proj_l (arrow Neg F) (inter (arrow Zero T)
                                                             $\,\,$(arrow Pos F)) is_0))
        by apply Eqproj_l.
      apply Eqsymm in H.
      apply Eqsymm in H0.
      eapply Eqtrans.
      apply H0.
      eapply Eqtrans.
      apply H.
      apply H1.
    - trivial.
  Defined.

End Test.
\end{coq}
The code of \coqe|is0test| has been generated by an Ltac script, the generated code is too huge to be humanly readable, as you can see in Figure \ref{pierce-coq}.
\begin{figure}[t!]
\begin{coq}
Copair Pos Neg F
     (Abst Pos F
        (fun x : OK Pos =>
         App Pos F
           (Proj_r (arrow Zero T) (arrow Pos F)
              (Proj_r (arrow Neg F) (inter (arrow Zero T) (arrow Pos F)) is_0)) x))
     (Abst Neg F
         (fun x : OK Neg => App Neg F (Proj_l (arrow Neg F)
         (inter (arrow Zero T) (arrow Pos F)) is_0) x)) Test
     (Eqabst Pos F Neg F
        (fun x : OK Pos =>
         App Pos F
           (Proj_r (arrow Zero T) (arrow Pos F)
              (Proj_r (arrow Neg F) (inter (arrow Zero T) (arrow Pos F)) is_0)) x)
         (fun x : OK Neg => App Neg F (Proj_l (arrow Neg F)
         (inter (arrow Zero T) (arrow Pos F)) is_0) x)
        (fun (x : OK Pos) (y : OK Neg) (pf : Eq Pos Neg x y) =>
         Eqapp Pos F Neg F
           (Proj_r (arrow Zero T) (arrow Pos F)
              (Proj_r (arrow Neg F) (inter (arrow Zero T) (arrow Pos F)) is_0)) x
           (Proj_l (arrow Neg F) (inter (arrow Zero T) (arrow Pos F)) is_0) y
           (Eqtrans (arrow Pos F) (inter (arrow Zero T) (arrow Pos F)) (arrow Neg F)
              (Proj_r (arrow Zero T) (arrow Pos F)
                 (Proj_r (arrow Neg F) (inter (arrow Zero T) (arrow Pos F)) is_0))
              (Proj_r (arrow Neg F) (inter (arrow Zero T) (arrow Pos F)) is_0)
              (Proj_l (arrow Neg F) (inter (arrow Zero T) (arrow Pos F)) is_0)
              (Eqsymm (inter (arrow Zero T) (arrow Pos F)) (arrow Pos F)
                 (Proj_r (arrow Neg F) (inter (arrow Zero T) (arrow Pos F)) is_0)
                 (Proj_r (arrow Zero T) (arrow Pos F)
                    (Proj_r (arrow Neg F) (inter (arrow Zero T) (arrow Pos F)) is_0))
                 (Eqproj_r (arrow Zero T) (arrow Pos F)
                    (Proj_r (arrow Neg F) (inter (arrow Zero T) (arrow Pos F)) is_0)))
              (Eqtrans (inter (arrow Zero T) (arrow Pos F))
                 (inter (arrow Neg F) (inter (arrow Zero T) (arrow Pos F))) (arrow Neg F)
                 (Proj_r (arrow Neg F) (inter (arrow Zero T) (arrow Pos F)) is_0) is_0
                 (Proj_l (arrow Neg F) (inter (arrow Zero T) (arrow Pos F)) is_0)
                 (Eqsymm (inter (arrow Neg F) (inter (arrow Zero T) (arrow Pos F)))
                    (inter (arrow Zero T) (arrow Pos F)) is_0
                    (Proj_r (arrow Neg F) (inter (arrow Zero T) (arrow Pos F)) is_0)
                    (Eqproj_r (arrow Neg F) (inter (arrow Zero T) (arrow Pos F)) is_0))
                 (Eqproj_l (arrow Neg F) (inter (arrow Zero T) (arrow Pos F)) is_0))) pf))
\end{coq}
  \vspace{-3mm}
\caption{LF encoding of the Pierce's code}\label{pierce-coq}
  \vspace{-3mm}
\end{figure}
\end{document}